\documentclass[aps,amsfonts,pra,twocolumn,showpacs]{revtex4-1}

\usepackage{subfigure}
\usepackage{textcomp}
\usepackage{graphicx}
\usepackage{amssymb}
\usepackage{amsmath}
\usepackage{bm}
\usepackage{amsmath, amsthm, amssymb}
\usepackage{dsfont}
\usepackage[colorlinks=true,linkcolor=blue,urlcolor=blue,citecolor=blue]{hyperref}
\usepackage{url}

\newcommand{\aah}{Aubry-Andr\'{e} }
\newcommand{\schrod}{Schr\"{o}dinger equation }
\newcommand{\ipr}{\mathrm{IPR}}

\def\beq{\begin{equation}}
\def\eeq{\end{equation}}
\def\bea{\begin{eqnarray}}
\def\eea{\end{eqnarray}}

\begin{document}

\title{Self-dual quasiperiodic systems with power-law hopping}

\author{Sarang Gopalakrishnan}

\affiliation{Department of Engineering Science and Physics, CUNY College of Staten Island, Staten Island, NY 10314}
\affiliation{Physics Program and Initiative for the Theoretical Sciences, The Graduate Center, CUNY, New York, NY 10016}

\begin{abstract}
We introduce and explore a family of self-dual models of single-particle motion in quasiperiodic potentials, with hopping amplitudes that fall off as a power law with exponent $p$. These models are generalizations of the familiar \aah model. For large enough $p$, their static properties are similar to those of the \aah model, although the low-frequency conductivity in the localized phase is sensitive to $p$.
For $p \alt 2.1$ the \aah localization transition splits into three transitions; two distinct intermediate regimes with both localized and delocalized states appear near the self-dual point of the \aah model.
In the intermediate regimes, the density of states is singular continuous in much of the spectrum, and is approximately self-similar: states form narrow energy bands, which are divided into yet narrower sub-bands; we find no clear sign of a mobility edge. 
When $p < 1$, localized states are not stable in random potentials; in the present model, however, tightly localized states are present for relatively large systems.
We discuss the frequency-dependence and strong sample-to-sample fluctuations of the low-frequency optical conductivity, although a suitably generalized version of Mott's law is recovered when the power-law is slowly decaying. We present evidence that many of these features persist in models that are away from self-duality.
\end{abstract}

\maketitle

\section{Introduction}

Quasiperiodic structures such as quasicrystals~\cite{qcbook} are intermediate between regular lattices and random potentials: they are not spatially periodic (so Bloch's theorem does not apply) but possess sharp Bragg peaks. The electronic states of quasiperiodic media resemble crystals in some respects (e.g., hard band-gaps~\cite{thouless83, kohmoto1983}) and random systems in others (e.g., Anderson localization~\cite{aa, sokoloff}). 
Since quasiperiodic optical potentials are straightforward to implement, many ultracold atomic experiments exploring Anderson localization~\cite{inguscio, inguscio2, schreiber2015, luschen2016, bordia2017, sanchez2010} (as well as related phenomena like many-body localization~\cite{nhreview, iyer_qp, schreiber2015, luschen2016, bordia2017} and the Bose glass phase~\cite{inguscio2, giamarchi_qp}) use quasiperiodic potentials instead of random ones. 
Although localization happens in both quasiperiodic and random systems~\cite{aa, azbel, suslov1982, wilkinson1984}, the localized phases and localization transitions are in some respects quite different~\cite{aaVand}. 
Quasiperiodic structures have much weaker large-scale fluctuations than random structures: they are ``hyperuniform''~\cite{steinhardt2017}. 
Therefore, one expects rare-region effects such as Lifshitz tail states~\cite{kmreview}, Griffiths-McCoy singularities~\cite{griffiths, mccoy}, and the like, to be suppressed in these systems. 
The suppression of large-scale potential fluctuations also has implications for critical phenomena, which tend to be less strongly affected by quasiperiodic modulation than by quenched randomness. For example, critical exponents in quasiperiodic systems are not subject to the Harris~\cite{harris} and CCFS~\cite{ccfs} bounds, but only to the weaker Luck bound~\cite{luck}: the one-dimensional localization transition in a quasiperiodic potential explicitly violates the Harris and CCFS bounds~\cite{sokoloff}, and potentially so does the many-body localization transition~\cite{ksh}. 
Understanding which aspects of localization are ``universal'' and which are specific to uncorrelated randomness is thus of both practical and conceptual interest.

\begin{figure*}
\begin{center}
\includegraphics[width = \textwidth]{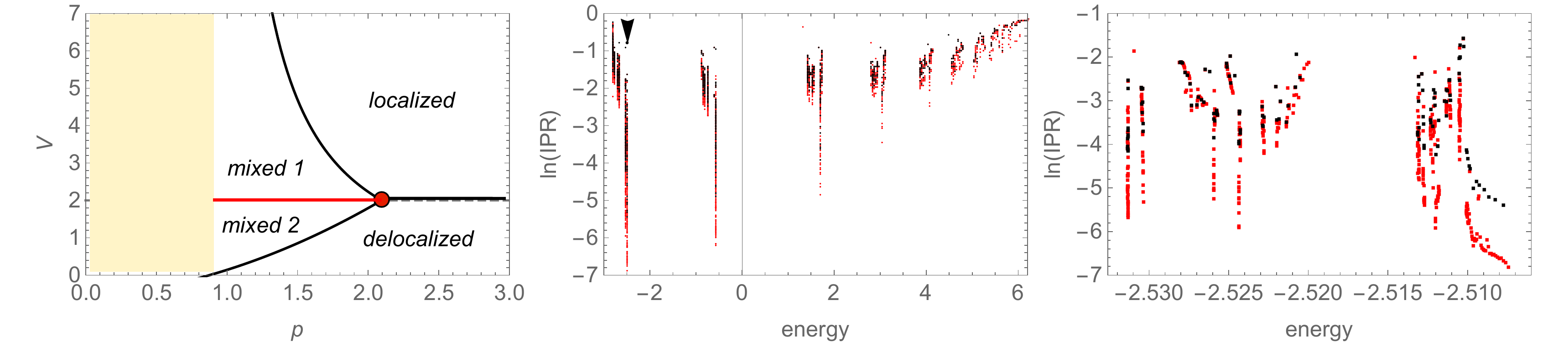}
\caption{Left: Phase diagram of the unfrustrated, self-dual power-law \aah model. For $p \alt 2.1$, intermediate mixed phases appear between the fully localized and fully delocalized phases. A phase transition at $V = 2$ remains for $p > 1$, separating the two mixed phases. For $p < 1$, a localization transition only occurs near the edge of the spectrum. Other panels show the structure of the mixed phase. Center: Logarithm of IPR vs. energy eigenvalue, for $p = 1.75, V = 3$, deep in the mixed phase. Red symbols are results for $L = 4000$; black symbols, for $L = 1000$. Upper panel shows all eigenstates, which are evidently clustered into very narrow bands. The band indicated with an arrow is pictured, magnified, in the right panel. Evidently there is considerable fine structure within this band, including apparently localized as well as clearly extended subbands.}
\label{pd}
\end{center}
\end{figure*}

Quasiperiodic potentials have been most extensively studied in one dimension in the tight-binding limit. In this limit the \schrod can be formulated as a tight-binding hopping model, the \aah model~\cite{aa}:

\beq\label{aamodel}
H_{\mathrm{AA}} = \sum_n (|n \rangle \langle n+1| + \mathrm{h.c.}) + V \cos(q n) |n \rangle \langle n|,
\eeq
where the period of the deeper lattice and the nearest-neighbor hopping strength have been set to unity; $n$ indexes sites of the lattice; and $q$ is a wavevector that is incommensurate with the underlying lattice periodicity. 
This model is not entirely generic, as it has a ``duality'' between the weakly and strongly disordered phases~\cite{aa, sokoloff}: the position-space \schrod equation at potential strength $V$ maps onto the momentum-space \schrod equation at strength $4/V$; correspondingly, the position-space eigenfunctions at strength $V$ are the Fourier transforms of those at strength $4/V$.  
This duality maps Anderson localized wavefunctions at large $V$ onto ballistically propagating waves at small $V$: the former are localized in position space; the latter in momentum space. 
At $V = 2$, the \schrod equation maps to itself, and the wavefunctions are statistically similar to their own Fourier transforms (and thus are delocalized in both position and momentum space). Transport at this point appears to be diffusive~\cite{varma2017, manas2017}. 
The entire spectrum of this model simultaneously changes its character at $V = 2$, so there is no mobility edge. The phase structure of the \aah model is the simplest consistent with the duality: for $V < 2$, all states are extended and ballistic, whereas for $V > 2$ all states are localized.
These features are specific to the self-dual model. Generic perturbations, though irrelevant at the critical point, will break self-duality and generically cause energy-dependent mobility edges to appear~\cite{biddle2010, biddle2011, sds3, sds4}.

\begin{figure}[tb]
\begin{center}
\includegraphics[width = 0.45\textwidth]{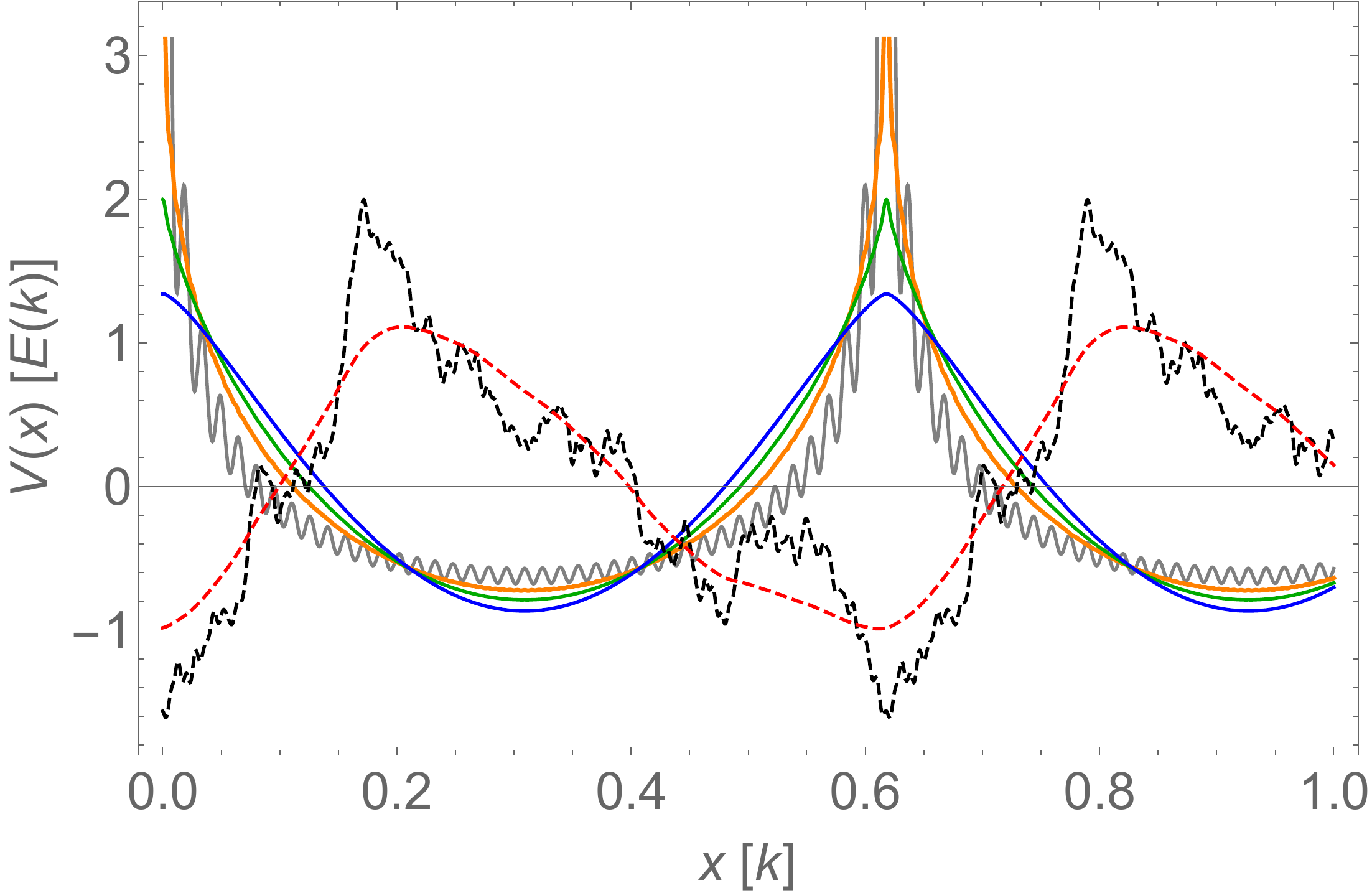}
\caption{Incommensurate potential as a function of position, $V(x)$ [or, equivalently, dispersion relation $E(k)$] for powers $p = 0.6, 1.2, 1.7, 2.5$ in the unfrustrated case (solid lines: respectively, gray, orange, green, and blue) and for $p = 1.2, 2.5$ in the highly frustrated case (dashed lines: respectively, black and red).} 
\label{disps}
\end{center}
\end{figure}

Note that the phase structure of the \aah model is simpler than self-duality mandates. In principle, there could be \emph{multiple} phase transitions---which must, of course, occur either at the self-dual point or at pairs of points $(V, 4/V)$. There could also be an intermediate self-dual phase encompassing the point $V = 2$. A natural way to open up these new possibilities is to add longer-range couplings to the \aah model~(Fig.~\ref{pd}). To this end, the present work explores a family of generalizations of the \aah model, in which the hopping falls off as a power-law in space, and the potential is altered so as to maintain self-duality under the simple \aah (``Fourier-transform'') mapping:

\bea\label{PLAAmodel}
H_{\mathrm{PQBM}} & = & \sum_{m \neq n}^N \left( \frac{e^{i \theta_m}}{|n - m|^p} |n\rangle \langle m| + \mathrm{h.c.} \right.  \\ & & \quad + \sum_{n}\sum_{m > 0}^N \left. \frac{V}{m^p} \cos[q m (n + \phi) + \theta_m] |n \rangle \langle n| \right) \nonumber
\eea
This extended family of models---which we term power-law quasiperiodic banded matrices (PQBMs), by analogy with power-law random banded matrices (PRBMs)~\cite{mirlin_prbm}---is specified by the parameters $(V, p, \{\theta_m\})$, as well as a phase $\phi$ that sets the origin of the quasiperiodic potential relative to the underlying lattice. Throughout this work we take $q = 2\pi \varphi$, where $\varphi$ is the Golden Ratio, and use open boundary conditions. The parameter $p$ governs both the power law falloff of the hopping and the shape of the quasiperiodic potential~\cite{agg, monthus2017}; duality locks these features to each other. The parameters $\{\theta_m \}$ determine how ``frustrated'' the kinetic energy is---or, equivalently, how ``rough'' the potential is (Fig.~\ref{disps}). 
In most of this work, we consider the simplest version of this model, in which $\theta_m = 0$ for all $m$. We term this the ``unfrustrated'' model. However, we shall also briefly explore a subset of the frustrated models, for which $\theta_m = \theta \cos m$. 
All of these models are similar for large $p$, but differ for $p \alt 3$.

For $p \leq 1$, the sums in Eq.~\eqref{PLAAmodel} have divergences; correspondingly, some single-particle states have extensive energies in the thermodynamic limit. In such long-range models, it is often necessary to compensate for this divergence by scaling the kinetic energy term as some power of $N$. In this self-dual model, however, we need only require that the kinetic and potential terms scale similarly with $N$. The choice of truncation in Eq.~\eqref{PLAAmodel} ensures this, regardless of $p$. 

We find that the critical behavior of the unfrustrated model changes abruptly at $p \approx 2.1$ (Fig.~\ref{pd}). For $p \agt 2.1$, the critical behavior is essentially unchanged from its short-range \aah limit. For $p \alt 2.1$, the delocalization transition no longer involves the entire spectrum at once; instead, delocalized states appear for $V > 2$, and (more surprisingly) \emph{localized} states appear for $V < 2$. In the ``mixed phase'' with both delocalized and localized eigenstates, the two types of states are energetically separated. However, there is no clear mobility edge. Rather, the eigenstates cluster into sub-bands, which in turn have considerable fine structure. Each individual sub-band, however, appears to be either entirely localized, entirely delocalized, or critical. 
In the regime where $1 < p \alt 2.1$, there are three phase transitions as a function of $V$: a pair of transitions at small $V$ [large $V$] at which localized [delocalized] states appear in the spectrum, and a transition at $V = 2$ at which these states ``swap'' (i.e., localized states become delocalized and vice versa). The ``swapping'' transition has different critical properties from the conventional \aah transition.
Finally, at $p = 1$, the fully localized and fully ballistic phases disappear, as does the phase transition at $V = 2$. Instead, the unfrustrated model has a single phase, with extremely sparse wavefunctions; at the available system sizes, we cannot establish whether these states remain localized in the thermodynamic limit. 
Frustration shifts the phase boundaries but does not qualitatively change the phase diagram for $p > 1$. For $p < 1$, the frustrated models behave much more simply than the unfrustrated one: the eigenstates in the frustrated models are delocalized in both real and momentum space for all $V$. 

This paper is organized as follows. Sec.~\ref{pert} discusses the expected behavior of the PQBM in various limits, at a heuristic level. Sec.~\ref{unfrust} explores the phase diagram of the unfrustrated PQBM, and Sec.~\ref{frust} for some specific frustrated PQBMs. Sec.~\ref{mobedge} explores the nature of the energy spectrum, as well as the energy-dependence of the properties of wavefunctions, in the mixed phase.
Sec.~\ref{conductivity} discusses, and presents numerical evidence for, features in the low-frequency optical conductivity in the localized phase of the PQBM. Sec.~\ref{conclusions} summarizes our results.

\section{Expectations}\label{pert}

This section addresses various perturbatively accessible limits of the PQBM. The central numerical results of this paper are \emph{not} perturbatively accessible, but (reassuringly) our numerical and perturbative analyses agree when both are valid.

\emph{Large-$p$ behavior}.---When $p$ is sufficiently large [$p \agt 2.5$ in the unfrustrated model], the longer-range and higher-harmonic terms are essentially weak perturbations. The dispersion relation (and by duality the potential) have high-derivative discontinuities, but these have no direct physical significance. In this regime the main impact of the power-law hopping is to endow localized wavefunctions with power-law tails. One can see this within perturbation theory starting from the deeply localized limit: a wavefunction centered at site $i$ has a typical amplitude $1/(V |i - j|^{p})$ at site $j$. Thus the localization length (or ``Lyapunov exponent'') is not defined for any finite $p$. Whether the potential is quasiperiodic or random is immaterial for this argument. The power-law tails do affect the optical conductivity (which we will discuss in Sec.~\ref{conductivity}); however, our main diagnostic, the inverse participation ratio (IPR),

\beq
\ipr(\psi) \equiv \frac{\sum_i |\langle i | \psi \rangle|^4}{\left( \sum_i |\langle i | \psi \rangle|^2 \right)^2}
\eeq
is relatively insensitive to power-law tails. 
The critical state of the \aah model is power-law correlated, so power laws that fall off sufficiently fast should leave critical wavefunctions unaffected. 

\emph{Intermediate-$p$ behavior: $1 < p \alt 2$}.---When $p \alt 2$, the dispersion of the unfrustrated model changes qualitatively. A cusp develops at $k_0 = 0$ for $p = 2$; for $1 < p < 2$, the dispersion around band-edge takes the form $E(k) \sim |k - k_0|^{p - 1}$; the density of states at $k_0$ vanishes at this point for $p < 2$. Ballistic propagation of states near $k_0$ is therefore robust, as the low density of states makes the system less susceptible to perturbations. By the same reasoning, localization should also be robust for states where the quasiperiodic potential is near its maximum value $V(x_0) = V(x = 0)$. 
The density of states halfway across the Brillouin zone, at $k = \pi$, increases: that part of the band flattens with decreasing $p$, so states there become more susceptible to perturbations. 

\begin{figure}[tb]
\begin{center}
\includegraphics[width = 0.45\textwidth]{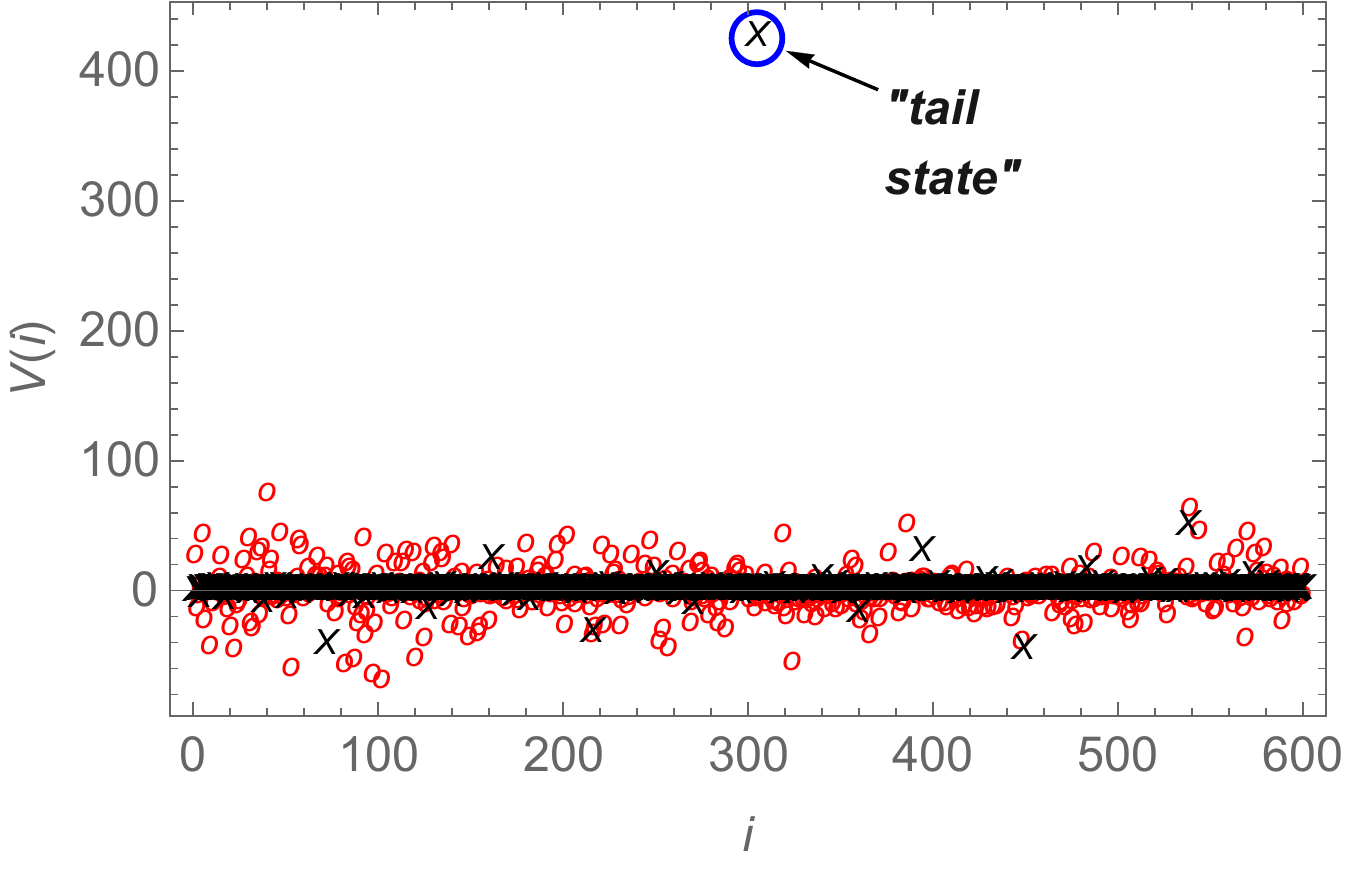}
\caption{Incommensurate potential vs. lattice site for $p = 0$ in the unfrustrated (black $x$'s) and highly frustrated (red $o$'s) models. In the unfrustrated model, most states (the ``bulk'') form a flat band, with a few outliers (``tail states'') that are very far away in energy. In the frustrated model, outliers are suppressed but the bulk has a less flat band.}
\label{smallp}
\end{center}
\end{figure}

\begin{figure}[tbp]
\begin{center}
\includegraphics[width=0.4\textwidth]{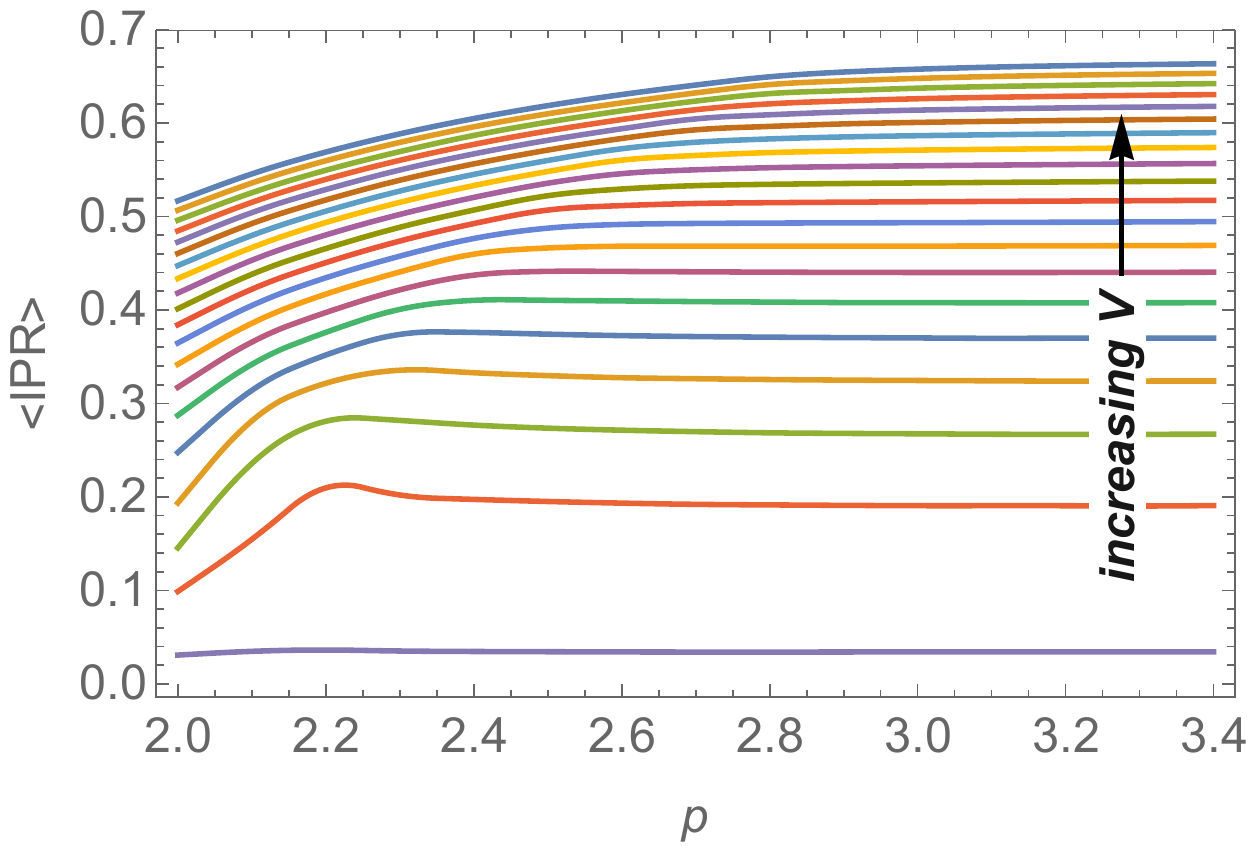}
\includegraphics[width=0.4\textwidth]{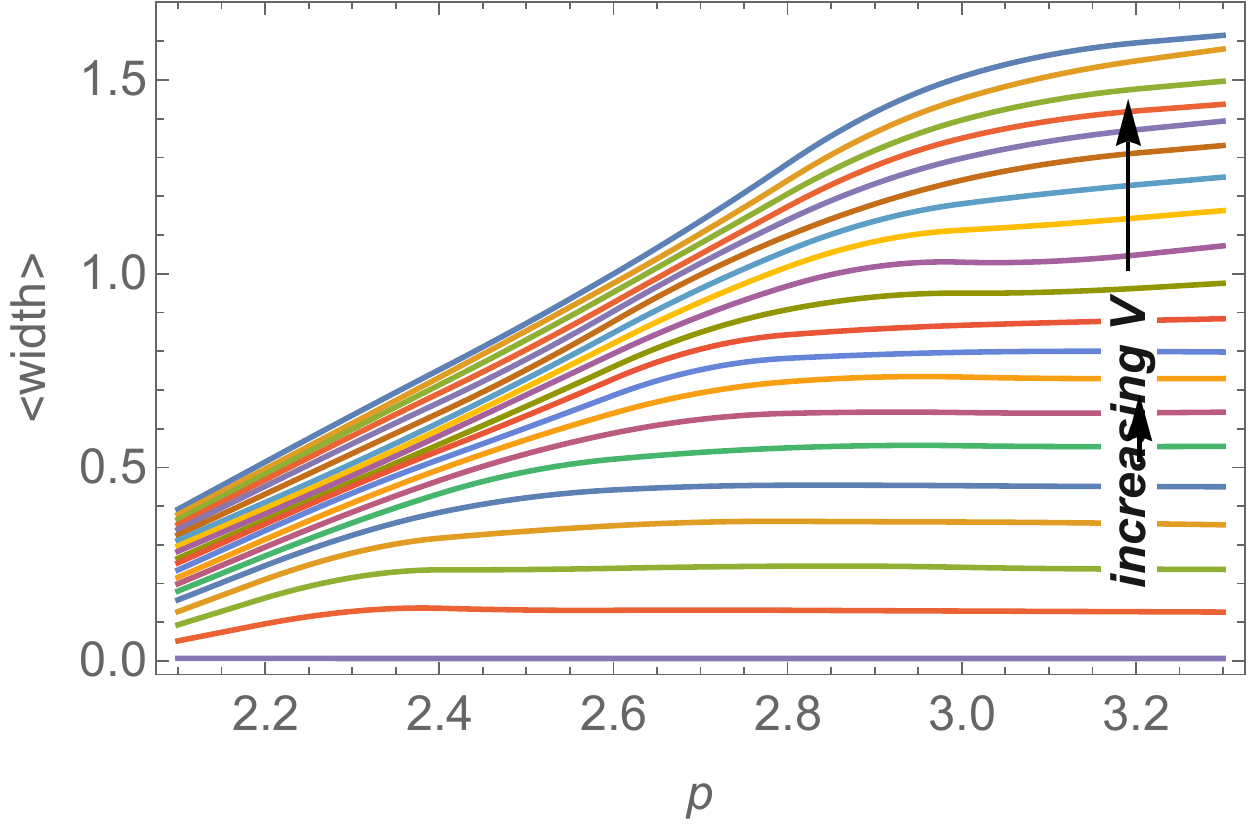}
\caption{Mean IPR and mean width as a function of $(V,p)$. Each trace is for a fixed $V$ as a function of $p$. $V$ ranges from 2 (bottom) to 4 (top). Deep in the localized phase, both quantities are sensitive to $p$; however, for $V$ near the transition, the $p$-dependence becomes flat except as $p \rightarrow 2$. Data are for $L = 1000$, and are averaged over ten values of the phase $\phi$.}
\label{collapse}
\end{center}
\end{figure}

\emph{Small-$p$ behavior: $p < 1$}.---
It is helpful to consider the all-to-all limit, $p = 0$. Let us first consider the unfrustrated case. Here, the single-particle dispersion is flat at zero energy for all states $k \neq 0$. The zero-momentum state has an energy that is extensive in system size, while the incommensurate potential is a set of delta functions (of strength $\propto N$). These delta-function spikes ``miss'' almost all lattice sites. The $k = 0$ state and the states (if any) localized at delta-function peaks are extensively far in energy from the bulk of the states, and can be regarded as having been projected out. The bulk of the states, in this model, are at precisely zero energy, regardless of $V$. In this limit, the bulk of the spectrum forms a flat band. 
A level crossing takes place at one edge of the spectrum when $V = 2$: the highest-energy state (in our sign convention) crosses from being a fully delocalized $k = 0$ state to being a single-site localized state. This level crossing can be regarded as a first-order thermodynamic phase transition; the excitation spectrum is blind to this transition.

Adding frustration changes this picture entirely. Consider, for example, the case where the quantities $\{\theta_m\}$ are chosen independently and randomly. While the unfrustrated model has a flat bulk band and a few faraway-in-energy tail states, the frustrated model has a random on-site potential as well as random long-range hopping, i.e., it is described by a conventional random matrix. The eigenstates in this model are delocalized in both real space and momentum space for all $V$. 

Away from the all-to-all limit, for $0 < p < 1$, these arguments do not directly apply. However, numerically we see a stark difference between the unfrustrated and frustrated models in this regime. In the frustrated model (Sec.~\ref{frust}), all states are delocalized in both position and momentum space, for all $V$, when $p < 1$. This is the same behavior as one sees in the PRBM~\cite{levitov_epl, mirlin_prbm, quito}. But in the unfrustrated model (Sec.~\ref{unfrust}), localized and delocalized states are interspersed irregularly throughout the spectrum for all $V$. However, states near the edge of the spectrum undergo a localization transition at $V = 2$. Thus, the long-range, $0 < p < 1$ regime appears to inherit the qualitative behavior of the corresponding all-to-all limit.


\section{Phase diagram of the unfrustrated model}\label{unfrust}

\subsection{Change of critical behavior}

\begin{figure}[tbp]
\begin{center}
\includegraphics[width=0.4\textwidth]{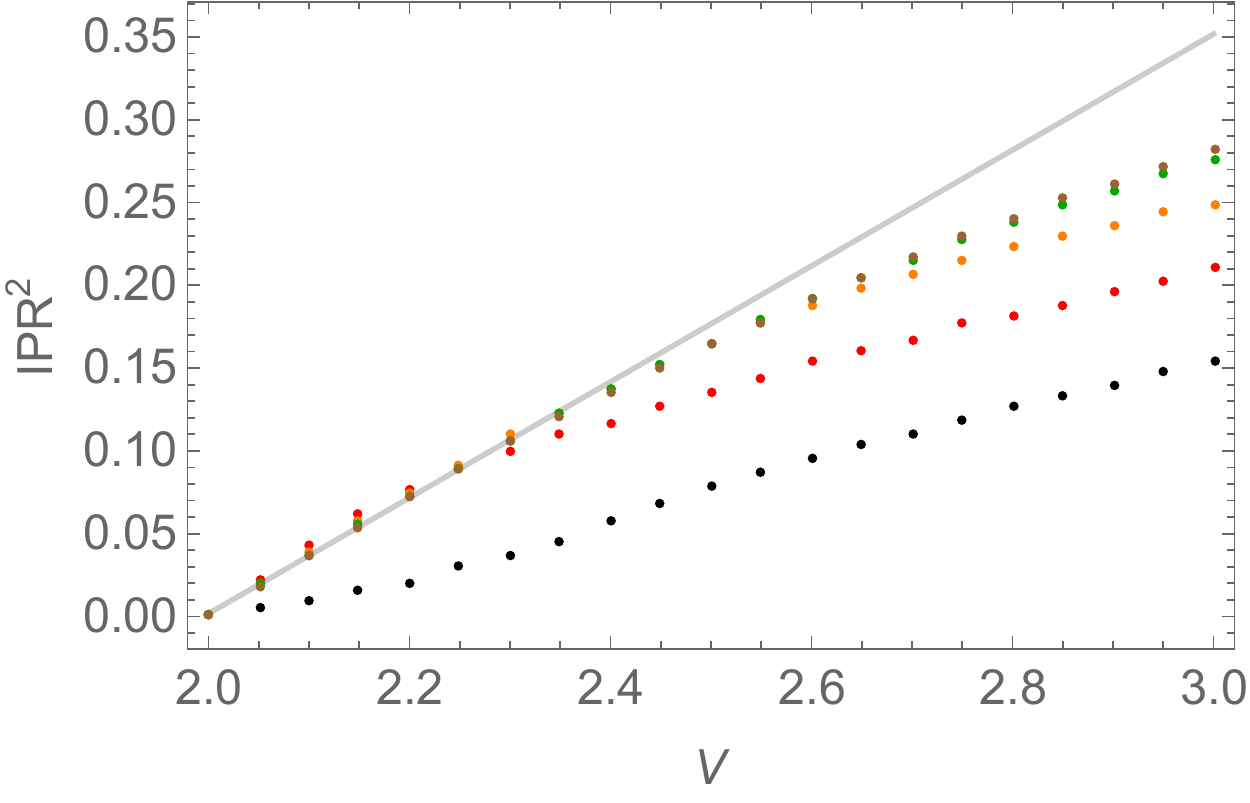}
\caption{Square of the IPR  as a function of $V$; $p$ ranges from 2 (lowest curve) to 3 (uppermost curve). The data collapse reasonably well, except for $p = 2$, for which the critical scaling is manifestly different.}
\label{collapse2}
\end{center}
\end{figure}

Rapidly decaying power-law hopping is relevant in the localized phase, but presumably irrelevant at the critical point, since the critical point is power-law correlated even in the \aah model. Thus, one expects that as $V \rightarrow 2$, physical quantities should become increasingly insensitive to $p$. This is indeed what we find, for both the mean IPR and the mean width of a wavefunction (Fig.~\ref{collapse}). (The width is defined as $w(\psi) \equiv \langle \psi | \hat{x}^2 | \psi \rangle - (\langle\psi | \hat{x} |\psi \rangle)^2$. Fig.~\ref{collapse2} shows a data collapse for the IPR at various $p$; there is general agreement for $p > 2$, as well as a strong deviation for $p = 2$.

\begin{figure}[tb]
\begin{center}
\includegraphics[width=0.45\textwidth]{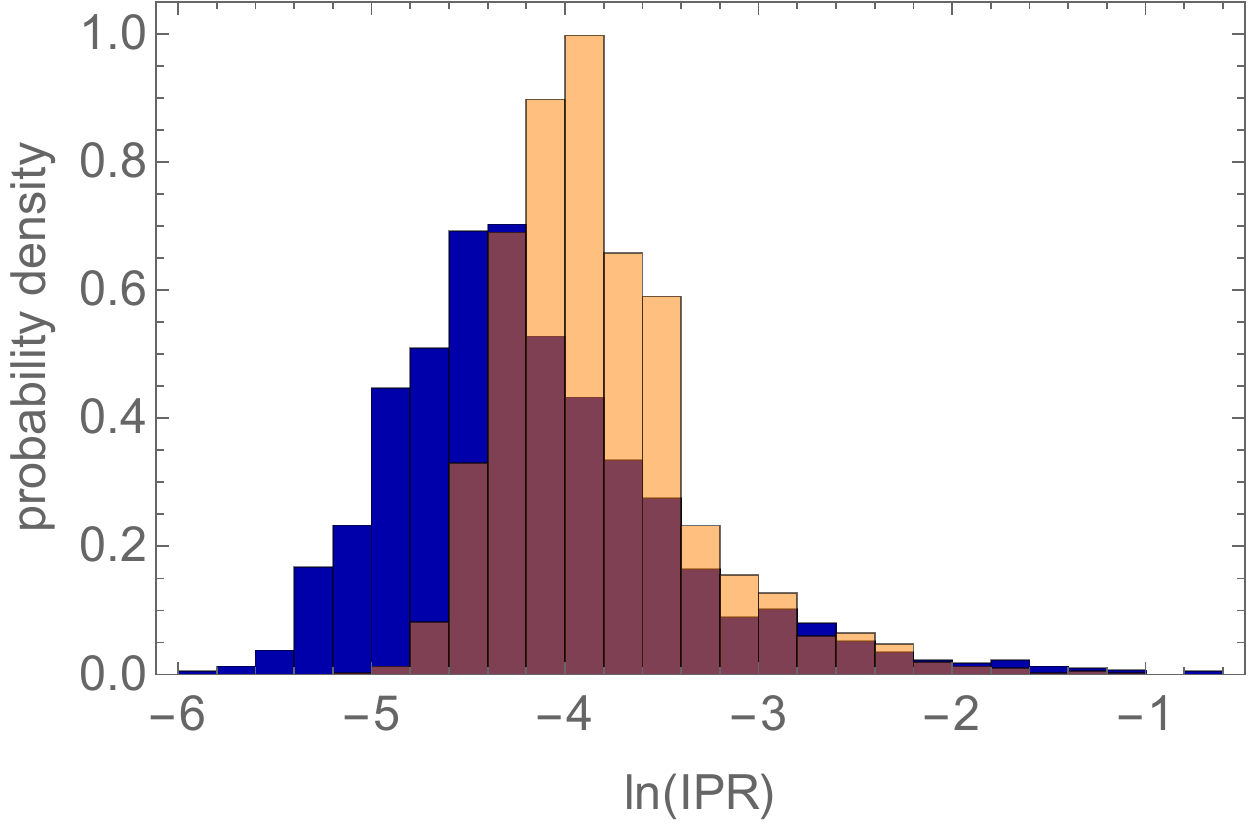}
\caption{Histogram of $\ln(\ipr)$ for $V = 2$, $p = 2$ (blue) and $p = 3$ (orange, translucent), for one particular sample with $L = 2000$. The $p = 2$ distribution is evidently wider.}
\label{iprex}
\end{center}
\end{figure}

\begin{figure}[tb]
\begin{center}
\includegraphics[width=0.45\textwidth]{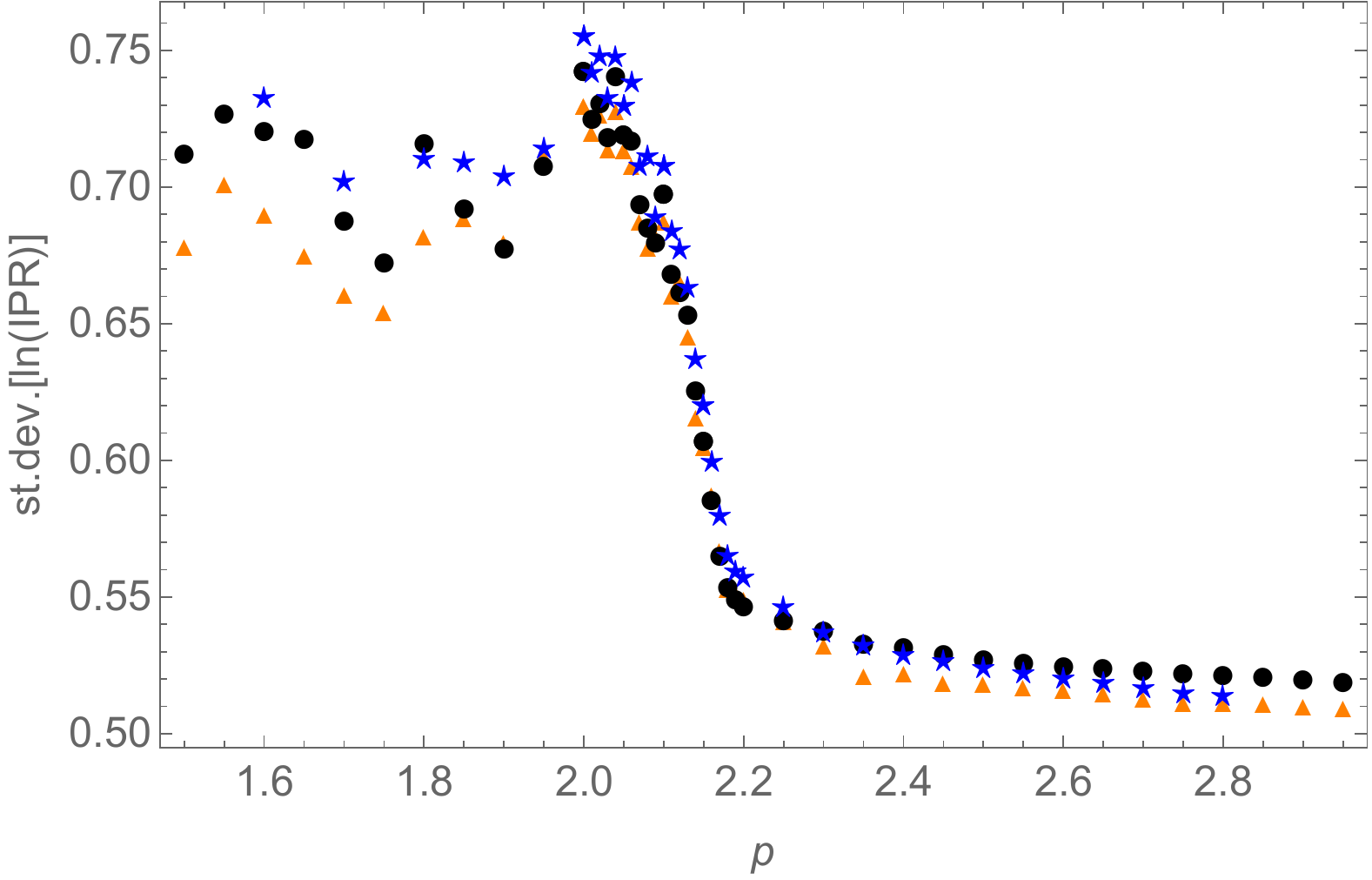}
\caption{Heterogeneity at $V = 2$ for various $p$. Red triangles are for $L = 2000$, black circles, $L = 3000$, and blue stars, $L = 4000$. Note the abrupt jump around $p = 2.1$.}
\label{lniprplot}
\end{center}
\end{figure}

We now turn to the properties of the critical state, $V = 2$, as $p$ is varied. From the histogram of IPRs (Fig.~\ref{iprex}), it is clear that the critical wavefunctions at $p = 2$ are more heterogeneous than at $p = 3$. To capture this property, we compute the standard deviation of the logarithmic IPR:

\beq\label{HET}
\mu \equiv \langle (\ln \mathcal{I}_\psi)^2 \rangle - ( \langle \ln \mathcal{I}_\psi \rangle)^2.
\eeq
We call this quantity the \emph{heterogeneity}. 
In the \aah model, the heterogeneity is size-independent in the localized phase, while it decreases slowly with system size in the ballistic phase; 
at the \aah critical point, the heterogeneity increases weakly with system size (App.~\ref{appB}).
We find numerically (Fig.~\ref{lniprplot}) that the heterogeneity at $V = 2$, at fixed system size, is largely $p$-independent for $p \agt 2.5$, but jumps sharply around $p = 2.1$. 

\subsection{Mixed phase for $p \alt 2.1$}

\begin{figure}[tb]
\begin{center}
\includegraphics[width=0.45\textwidth]{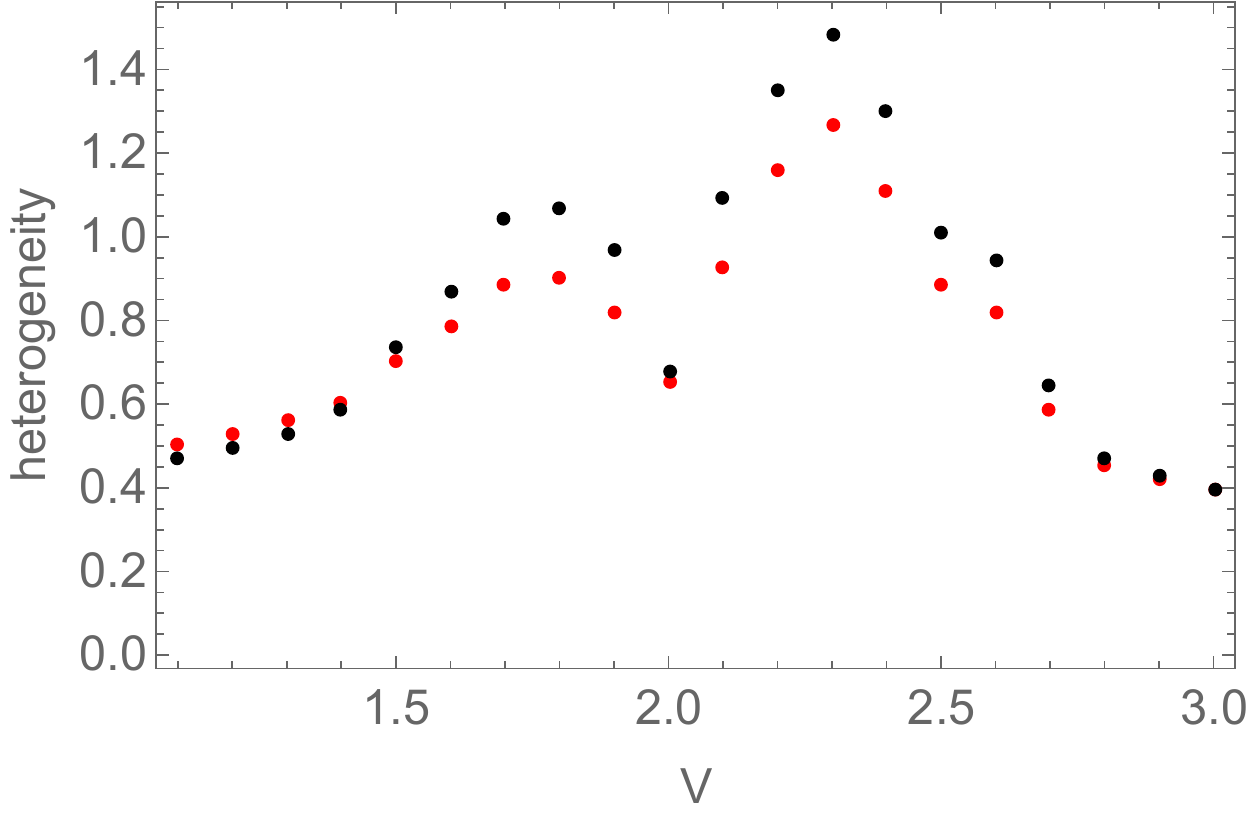}
\includegraphics[width=0.45\textwidth]{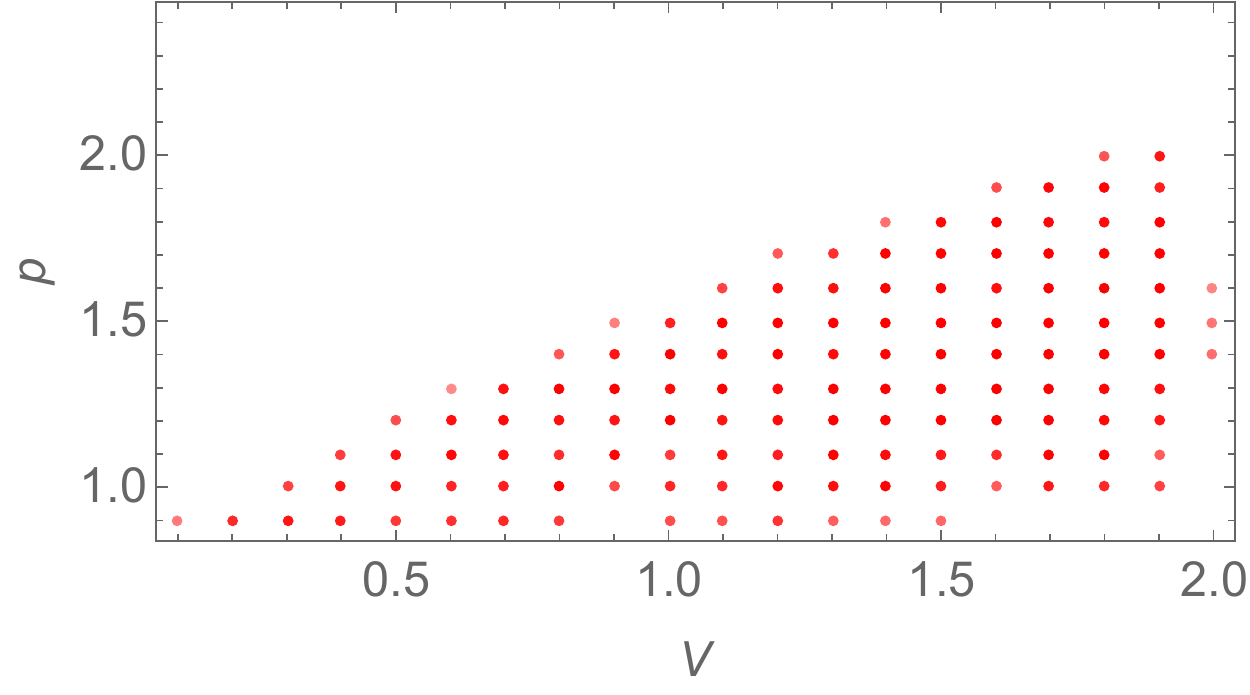}
\caption{Upper panel: heterogeneity as a function of $V$ for $p = 1.9$, for system size $L = 1000$ (red) and $L = 2000$ (black). In the mixed phase and only there, the heterogeneity \emph{increases} with system size. Lower panel: numerically extracted phase diagram, for $V < 2, 1 < p \alt 2.5$. Points marked in red are in the mixed phase. The heterogeneity decreases with system size in the fully delocalized phase and increases in the mixed phase; this flow reversal allows us to identify a transition point. The corresponding transition for $V > 2$ follows from duality.}
\label{hetp19}
\end{center}
\end{figure}

To understand what causes the critical state to change abruptly at $p = 2.1$, it is helpful to study the points away from criticality for $p \alt 2.1$. Fig.~\ref{hetp19} shows the behavior of the heterogeneity as a function of system size for $p = 1.9$. There are two striking features here: first, there is a regime on either side of the self-dual point in which the heterogeneity increases strongly with system size. This behavior is what one would expect from a phase with coexisting localized and delocalized states (App.~\ref{appB}). Second, there is a sharp dip in the heterogeneity at $V = 2$. This suggests that a phase transition at $V = 2$ survives, but separates two distinct ``mixed'' phases, in which some states are localized and others are extended. There is a mostly delocalized phase at $V < 2$, and a mostly localized phase for $p > 2$. At $V = 2$, the localized and delocalized states ``swap.'' As $p$ decreases, the mixed phase occupies a growing part of the phase diagram, until (apparently) the fully localized and fully delocalized phases vanish when $p \approx 1$ (Figs.~\ref{pd},~\ref{hetp19}). 
More details on the mixed phase, including evidence for localized states for $V < 2$, is presented in App.~\ref{casestudies}.

\subsection{Long-range potentials: $p < 1$}

\begin{figure}[tb]
\begin{center}
\includegraphics[width = 0.45\textwidth]{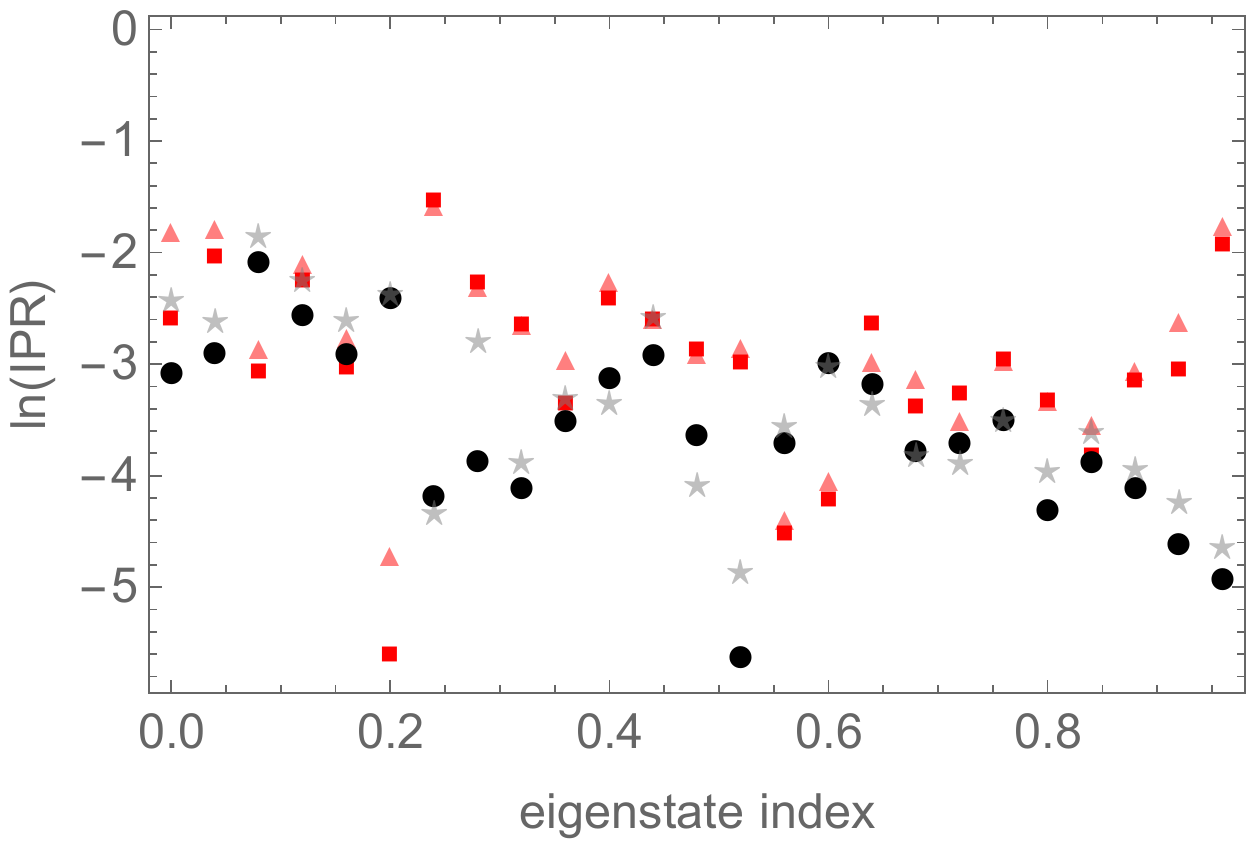}
\caption{Logarithm of IPR vs. eigenstate index for $p = 0.75$. $V = 4$ (red) and its dual point $V = 1$ (black/gray). Bold symbols are for $L = 4000$; faint symbols, $L = 2000$. States are binned into 25 bins and $\ln$(IPR) is averaged for each bin.}
\label{p075}
\end{center}
\end{figure}

As $p \rightarrow 1$ from above, the fully localized and ballistic phases vanish. Instead, for $p < 1$ the distribution of IPRs appears to be strongly heterogeneous for all $V$. Fig.~\ref{p075} shows the partially coarse-grained behavior of the IPR across the spectrum, for the dual points $V = 1$ and $V = 4$. For both values of $V$, much of the spectrum seems delocalized (i.e., the IPR decreases with system size) but parts of the spectrum are evidently localized (i.e., the IPR seems size-independent, or even increases slightly with system size~\footnote{For $p < 1$ the potential changes appreciably with system size, so this behavior is not particularly unexpected.}). In much of the spectrum, localized and delocalized regions are intermixed without any apparent pattern---a feature that is also present at the self-dual point $V = 2$. However, states near one end of the spectrum (corresponding to the singularity of the dispersion [Fig.~\ref{disps}], and thus to an anomalously small density of states) are clearly delocalized for $V < 2$ and clearly localized for $V > 2$. This is consistent with our expectation that this part of the spectrum should be ``robust'' in either phase.

The persistence of localized states, even for $V = 1$ and relatively large systems (up to $L = 6000$), is unexpected, given the absence of localized states in the PRBM (and, as we shall see below, in the frustrated PQBM). We propose the following heuristic picture of these states. Let us take $V \gg 1$. The potential is nearly flat at most values, separated by steep, quasiperiodically separated walls (Fig.~\ref{disps}). Within each flat domain, we can solve for the local band structure, in terms of a tight-binding model; since the dispersion is also nearly flat, this tight-binding model yields a discrete set of localized states within each domain. The domains are irregular in length (owing to the quasiperiodic distribution of spikes) and cannot easily resonate with one another. This accounts for the apparent localization even for relatively large system sizes; however, establishing whether any states remain localized in the thermodynamic limit in this regime requires a more careful analysis (including numerics on much larger system sizes) and we defer it to future work.

\section{Phase diagram of frustrated PQBMs} \label{frust}

\begin{figure}[tb]
\begin{center}
\includegraphics[width = 0.45\textwidth]{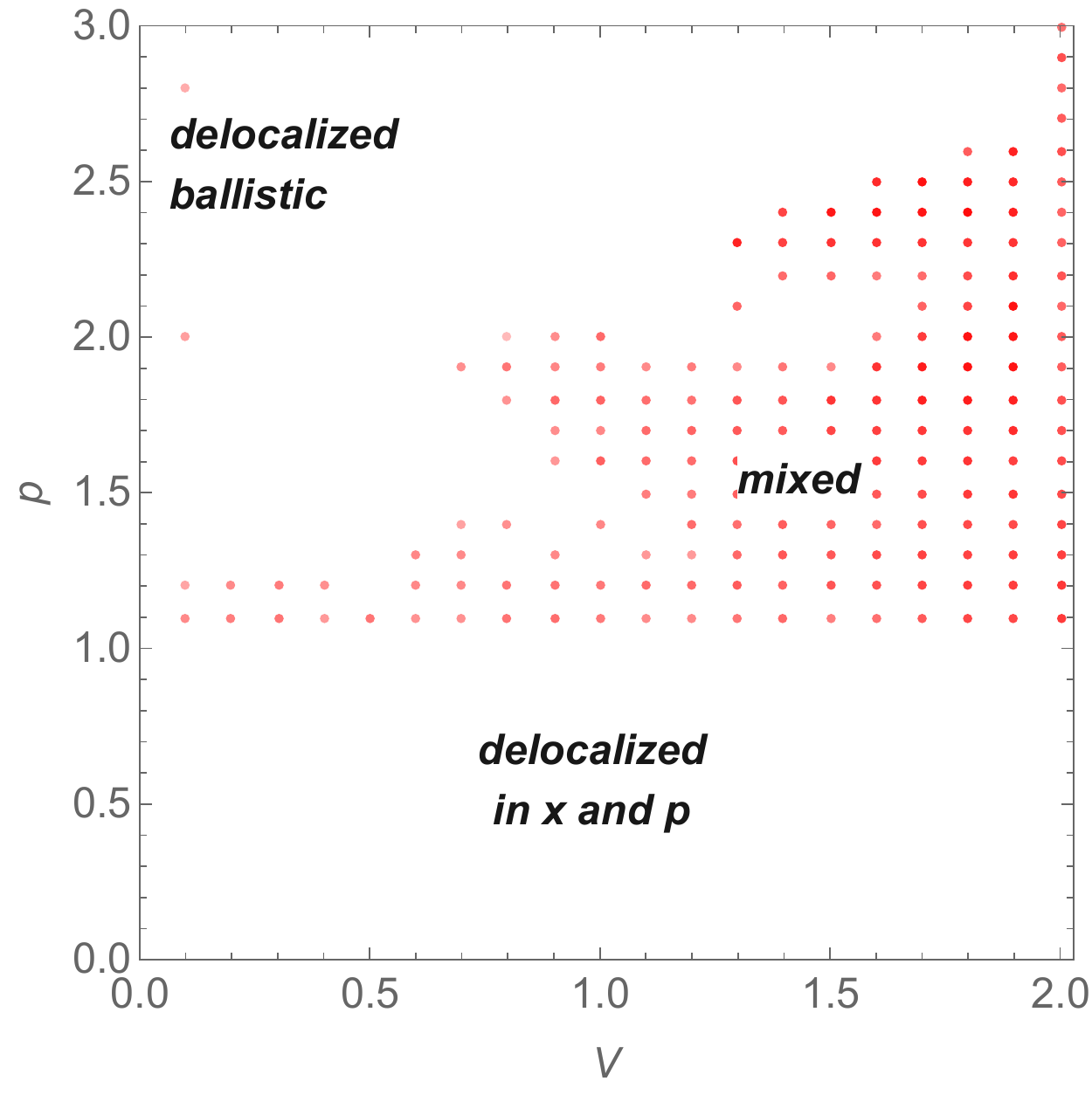}
\caption{Phase diagram of the frustrated PQBM for the specific choice $\theta_m = 100 \cos m$. Dots indicate the mixed phase, which extends to larger $p$ than in the unfrustrated case. For $p < 1$ essentially all states appear to delocalize.}
\label{frustpd}
\end{center}
\end{figure}

\begin{figure}[tbp]
\begin{center}
\includegraphics[width = 0.45\textwidth]{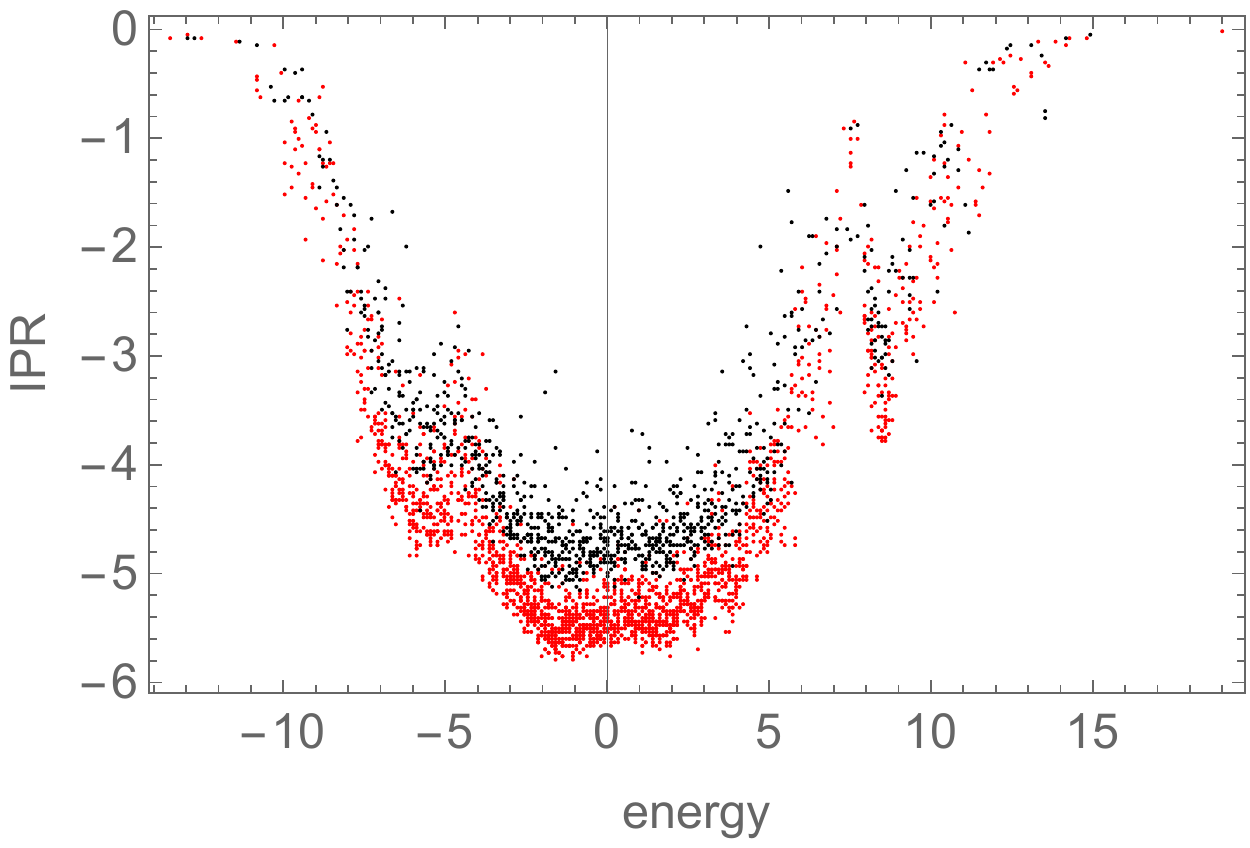}
\caption{Logarithm of IPR vs. energy eigenvalue for $V = 4, p = 0.8$, in the frustrated PQBM. Note the much simpler structure here than in Fig.~\ref{p075}.}
\label{frustLR}
\end{center}
\end{figure}

We now turn to the frustrated PQBMs, in which the phases $\{\theta_m\}$ are nontrivial. For concreteness we take $\theta_m = \theta \cos m$, and focus on a large value $\theta = 100$. (We investigated other values of $\theta$ and these are qualitatively similar.) This choice of $\{\theta_m \}$ gives rise to a potential that, at small $p$, fluctuates rapidly from site to site, but has few large-scale fluctuations: whereas random $\{ \theta_m \}$ would give $\sim \sqrt{N}$ site-to-site fluctuations, for our choice of $\{ \theta_m \}$ these fluctuations grow much more weakly, if at all. 
Nevertheless, the potential is very rough for $p \alt 2$ (Fig.~\ref{disps}). For $p > 1$, the phases of this model are similar to those of the unfrustrated PQBM. The phase boundaries are, however, somewhat different for large $\theta$: the mixed phase sets in at higher values of $p$, and has a somewhat different shape (Fig.~\ref{frustpd}). For $p < 1$, however, the frustrated models behave more conventionally than the unfrustrated one (Fig.~\ref{frustLR}). As with the PRBM~\cite{levitov_epl, mirlin_prbm}, resonances prevent localization, and all states are apparently delocalized.

\section{Spectral structure} \label{mobedge}


It would seem natural for a Hamiltonian with energy-separated localized and extended eigenstates to have a mobility edge. However, the spectral structure of the mixed phase is more subtle (Fig.~\ref{pd}): the eigenstates cluster into a large number of very narrow bands, each of which is further divided into many sub-bands, some of which have further internal substructure. Clearly extended and clearly localized sub-bands are found next to one another, as in the right panel of Fig.~\ref{pd}: we have not been able to find any clear pattern to their alternation. Moreover, we have not found any cases in which the same sub-band hosts both localized and extended states. Rather, each sub-band appears to be entirely localized, delocalized, or critical; in Fig.~\ref{pd}, the states that are clearly localized or delocalized correspond to bands that are not purely vertical; the vertical bands appear to host critical states. These remarks apply to the spectrum away from the edges, particularly the upper edge (where the density of states vanishes). Close to the upper edge, the spectrum is regular with no obvious signs of fractality, and both the density of states and the IPR behave much as one would expect in the short-range \aah model away from $V = 2$.

\begin{figure}[tbp]
\begin{center}
\includegraphics[width = 0.45\textwidth]{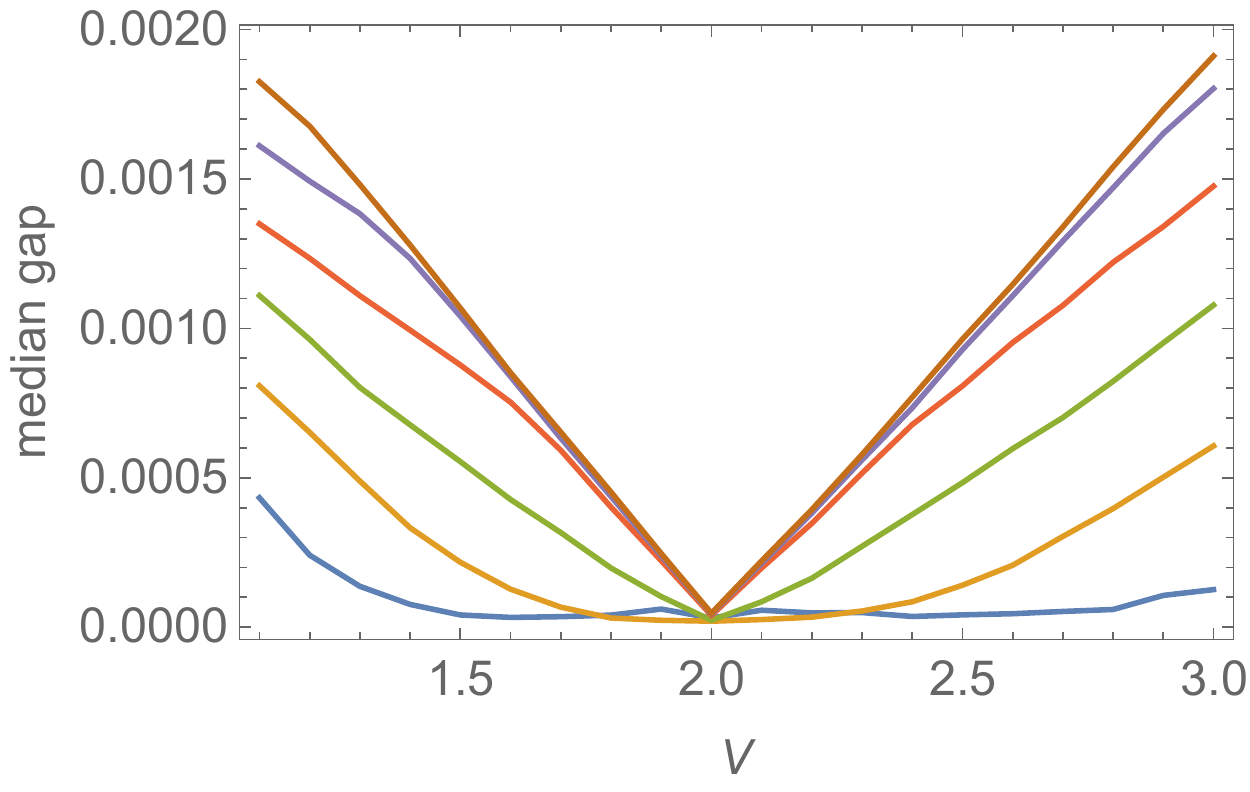}
\includegraphics[width = 0.45\textwidth]{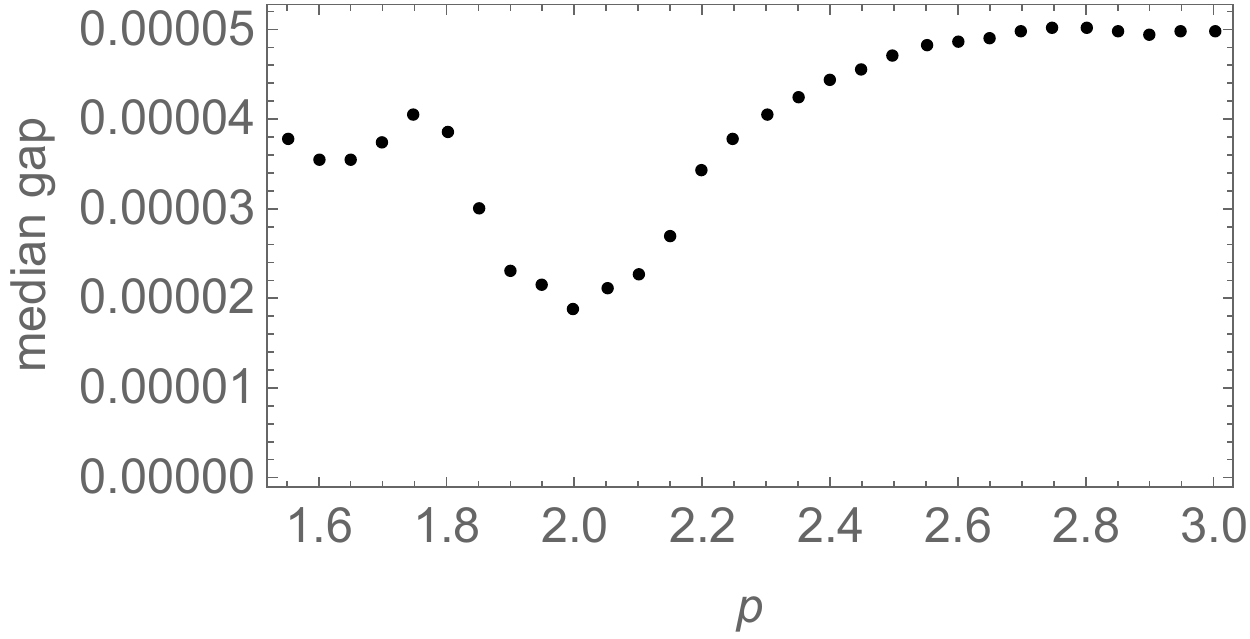}
\caption{Upper panel: median gap vs. $V$ for $p = 1.7, 1.9, 2.1, 2.3, 2.5, 2.7$ (lowest curve to uppermost). For $p > 2.1$ the curves behave qualitatively as in the \aah model. In the mixed phase at $p < 2.1$ the median gap is parametrically smaller than the mean gap. Lower panel: median gap at $V = 2$ for various $p$.}
\label{mediangap}
\end{center}
\end{figure}

Fig.~\ref{mediangap} quantifies this observation that the mixed state has very narrow bands by computing the median gap as a function of $V$ and $p$. For $p \agt 2.1$ the behavior is similar to that in the \aah model: the median gap is of order the mean gap except at the critical point; it decreases linearly as $V \rightarrow 2$. For $p \alt 2.1$ the critical behavior of the median gap changes; this gap appears to remain critical throughout the mixed phase (as we might expect from inspecting the band structure of Fig.~\ref{pd}). The lower panel shows the median gap at the critical point as $p$ is varied: it abruptly decreases at $p \approx 2.1$, tracking the other diagnostics of this transition, but then increases again as one moves to smaller $p$.


\section{Optical conductivity in the localized phase} \label{conductivity}

\begin{figure}[tb]
\begin{center}
\includegraphics[width = 0.45\textwidth]{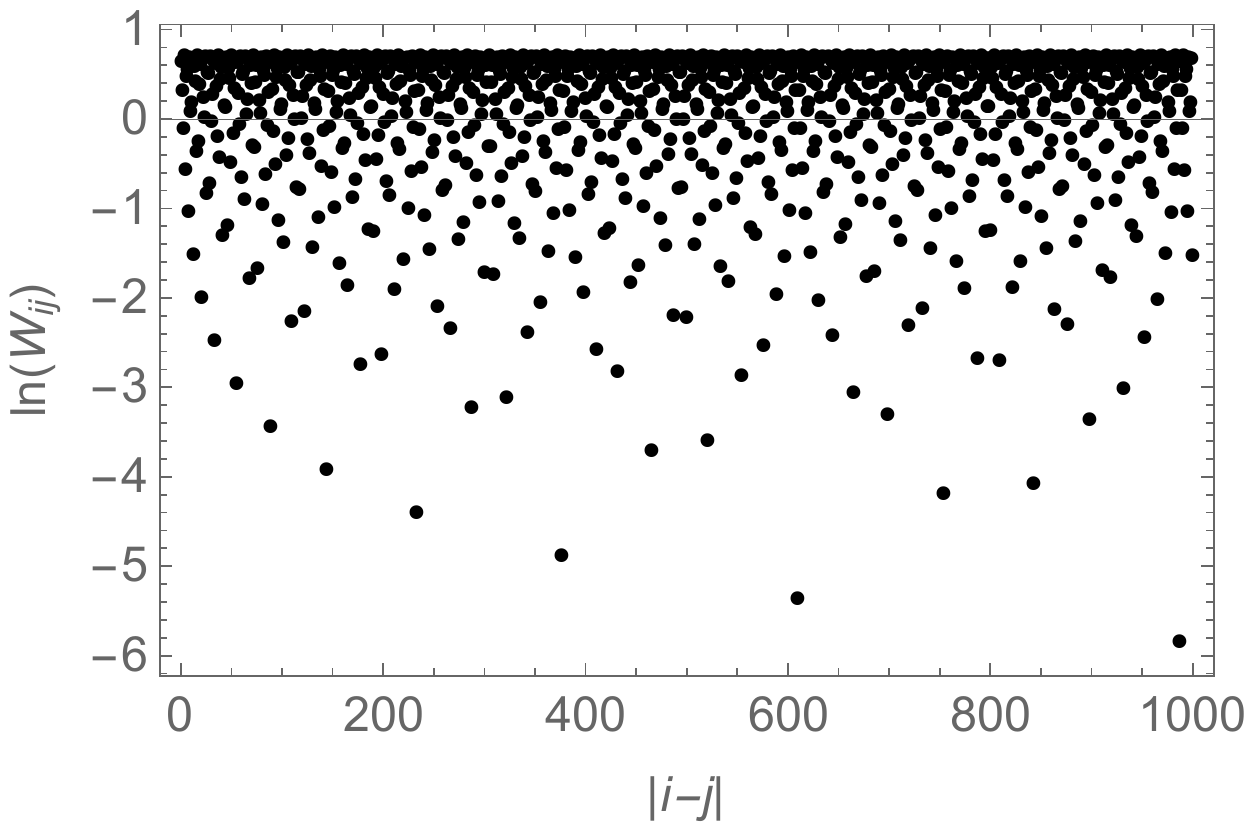}
\caption{Effective bandwidth $W_{ij}$ (see main text) as a function of separation $|i - j|$ for the \aah model with a Golden Rule potential, as in the main text. For a random system $W_{ij}$ is independent of $|i - j|$.}
\label{denoms}
\end{center}
\end{figure}

The nature of optical conductivity in localized quasiperiodic systems---even in the short-range limit---has seen relatively little discussion (but see Refs.~\cite{pgf, lahiri}); before we turn to power laws, we first discuss the more general problem. The theory for disordered systems in one dimension was established by Mott~\cite{mott1968} and Berezinskii~\cite{berez}. Mott argued that low-frequency conductivity deep in the localized phase is dominated by resonant pairs, i.e., eigenstates that live on two sites. To find the conductivity at $\omega$, therefore, one must estimate the density of resonant pairs with splitting $\sim \omega$, and therefore at separation $L_\omega \sim \xi \log(1/\omega)$. In random systems, a perturbative estimate in the hopping is straightforward, since the on-site energies at different sites are uncorrelated. In quasiperiodic systems, on the other hand, the energies at site $x$ and $x + L_\omega$ are obviously correlated; thus, one cannot directly apply resonance-counting arguments from the random case. 

Indeed, it is not clear whether there is a universal answer for quasiperiodic systems, as the statistics of energy differences at a distance $L$ are model-dependent. In the model of Ref.~\cite{pgf}, the quasiperiodic potential has the form $\tan x$; for this potential there are strictly \emph{no} resonances at most scales. Thus the optical conductivity in the model of Ref.~\cite{pgf} has a hard gap. The \aah model is different: the energy denominator between two sites at a distance $L$ apart is given by $V |\cos[q(n + \phi)] - \cos[q(n+L + \phi)]|$. A thermodynamically large system samples all values of $\phi$, and evidently \emph{some} choices of $\phi$ make the denominator vanish. Thus, resonances do exist in this model at all scales. Nevertheless, Mott's law is modified because the \emph{density} of resonances at distance $L$ is sensitive to $L$. 
We define the ``bandwidth'' $W_{ij} = \max_\phi\left\{ V |\cos[q(n + \phi)] - \cos[q(n+L + \phi)]| \right\}$; the logarithm of this quantity is plotted in Fig.~\ref{denoms}.
Evidently there are special spacings for which the bandwidth is anomalously small, so the density of resonances is anomalously high. The conductivity should spike when $L_\omega$ hits one of these anomalous distances. Presumably these spikes grow increasingly close to each other in frequency as $\omega \rightarrow 0$, so there is no well-defined frequency regime in which Mott's law applies. The detailed behavior of the optical conductivity in the \aah model is, however, outside the scope of the present work.

\begin{figure}[tb]
\begin{center}
\includegraphics[width=0.45\textwidth]{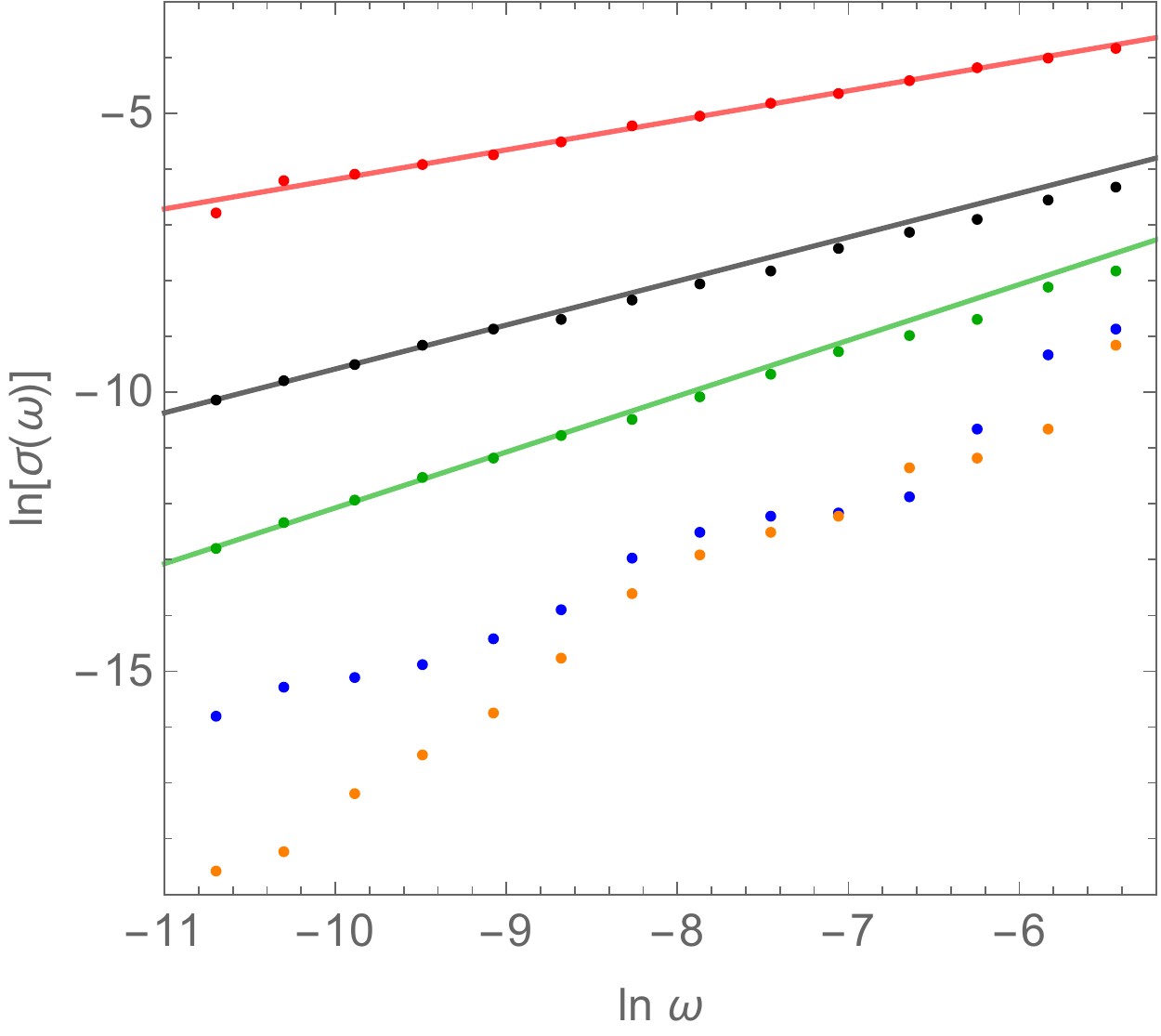}
\caption{Low-frequency conductivity (on a logarithmic scale) for $p = 2, 2.5, 3, 4, 8$ (from top to bottom), for $V = 4$ in the unfrustrated PQBM. These data are for $L = 1000$, averaged over 2000 values of the phase $\phi$. Linear fits on the log-log scale (solid lines on the plot; fits are to the seven lowest-frequency points) yield power laws with exponents $\alpha = 0.52, 0.79, 1.00$ for $p = 2, 2.5, 3$ respectively. These numerical results agree well with the theoretically predicted exponents~\eqref{sigmaomega} $\alpha = 0.5, 0.8, 1$. For larger powers, the frequency-dependence is not power-law in the accessible frequency range.}
\label{condfig}
\end{center}
\end{figure}

Power-law hopping modifies the statistics of resonances and the a.c. conductivity in various ways. First, even in random systems, Mott's argument~\cite{mott1968} is modified: instead of the standard expression $\sigma(\omega) \sim \omega^2 \log^2 \omega$, we expect 

\beq\label{sigmaomega}
\sigma(\omega) \sim \omega^{2 - 3/p}
\eeq
for power law $p$~\footnote{This simple result is presumably known, but I have not been able to find a reference for it.}. The Mott conductivity diverges at low frequencies for $p \leq 1.5$; this is related to the divergence of the wavefunction width $\langle \psi | \hat{x}^2 | \psi \rangle -(\psi | \hat{x} | \psi \rangle)^2$ in that regime~\cite{mirlin_prbm}. The argument for Eq.~\eqref{sigmaomega} is as follows. In systems with power-law hopping, the scale $L_\omega \sim \omega^{-1/p}$. Moreover, the spatial range over which one can find resonances at frequency $\omega$ changes. In general, this is the spatial range over which the tunneling amplitude is of order $\omega$ (e.g., between $\omega/e$ and $\omega$). For short-range systems, this range is simply a localization length. However, for systems with power-law hopping, this range is also modified to $\omega^{-1/p}$. 

An important qualitative implication of this for the PQBM is that as the power law decreases, the low-frequency behavior of $\sigma(\omega)$ should become increasingly smooth. This is indeed what we find numerically (App.~\ref{appC}). 
%
Fig.~\ref{condfig} presents numerical results for the optical conductivity of the unfrustrated PQBM, deep in the localized phase ($V = 4$) for various power laws. These data are evaluated in the $T \rightarrow \infty$ limit, thus they correspond to the high-temperature (or equivalently random-initial-state) settings that are typical in cold-atom experiments~\cite{sanchez2010}. Like the other observables we considered, the overall level of the conductivity increases sharply around $p = 2$. For power laws $2 \alt p \alt 3$, the conductivity fits well to the form~\eqref{sigmaomega} [Fig.~\ref{condfig}]; the exponents match well with the theoretical predictions. For larger $p$, the conductivity fluctuates strongly as a function of frequency (as one might expect from the discussion above), and no simple power-law fit captures the low-frequency data.

Unlike the static quantities we looked at, the conductivity remains sensitive to $p$ even for relatively large $p$. Reliably extracting the exponent for large $p$, however, requires more intensive numerical work. We remark that a striking feature in the numerics is that even for relatively large system sizes $L = 2000$ and well localized wavefunctions, the sample-to-sample fluctuations are severe (App.~\ref{appC}).

\section{Conclusions and outlook} \label{conclusions}

We have introduced and explored the phase diagram of a family of self-dual one-dimensional systems with quasiperiodic potentials and power-law hopping amplitudes. We have found that even when the power law is relatively rapidly decaying ($1 < p \alt 2.1$ for the unfrustrated model, $1 < p \alt 3$ for the frustrated models), a mixed phase appears, in which some eigenstates are localized and others are delocalized. There is a multicritical point $V_c = 2, p_c \approx 2.1$ in the $(V,p)$ plane at which the localization transition first splits; the critical exponents of this point remain to be explored. 
For $p < p_c$, surprisingly, \emph{localized} states emerge at weak disorder in the self-dual model. Moreover, instead of a single mixed phase with a smoothly changing fraction of localized and delocalized states, we find two mixed phases separated by a critical point at $V = 2$. This critical point appears to be in a different universality class from the \aah critical point, with critical wavefunctions that are more heterogeneous than those at the \aah critical point. 
In the mixed phase, the eigenstates are apparently squeezed into a measure-zero subset of the many-body bandwidth; the typical level spacing correspondingly vanishes as $L^{-\gamma}, \gamma > 1$. 
The properties of the mixed phase(s), the self-dual transition between them, and the transitions at which ``wrong-phase'' eigenstates first appear in the spectrum, are interesting topics for future study. So are transport and nonequilibrium dynamics~\cite{igloi2013} in the mixed phase. The extent to which our results on the optical conductivity extend to other quasiperiodic hopping models, such as that of Ref.~\cite{pgf}, is also an interesting topic: based on our arguments, we might expect weak power laws to give rise to a nonzero low-frequency conductivity in that model.

The PQBM model discussed here was motivated by theoretical considerations; nevertheless, models analogous to it can be experimentally realized, e.g., in ion-trap experiments~\cite{islam2013}, since the potential can be shaped essentially at will using spatial light modulators~\cite{mattp2008, *zoran2013}. A drawback is, however, that at the achievable system sizes ($L \approx 20$) the signatures might not be especially sharp. An alternative method for directly dialing in the desired dispersion relations was recently demonstrated~\cite{bryce}, and might prove more scalable. In practice, self-duality will also presumably be broken. We expect the phases in the PQBM phase diagram (Fig.~\ref{pd}) to survive weak duality-breaking perturbations; we find numerically that many of the features are robust even in the limit where, e.g., the hopping is power-law and the potential is monochromatic, or vice versa~\cite{monthus2017} (see App.~\ref{appD}). Even in that limit, long-range hopping gives rise to a band with flat regions, and states in these flat regions are susceptible to localization at weak disorder.

Because rare-region effects are suppressed in quasiperiodic systems, they can potentially exhibit many-body localization with power-law interactions~\cite{burin1998, dipole_mbl} even if rare regions preclude many-body localization in random systems with power-law interactions~\cite{drh, dri, luitz2017}. Adding interactions to the PQBM---or to related power-law spin models that have an Ising duality on top of the Fourier-transform duality~\cite{chandran2017}---offers a natural way to explore these questions.

\section{Acknowledgments}

\noindent The author is grateful to A. Chandran, B. Gadway, D. Huse, V. Khemani, V. Oganesyan, and V. Varma for enlightening discussions, and to V. Varma for helpful comments on the manuscript.

\appendix

\section{Localized states for $V < 2$}\label{casestudies}

\begin{figure}[htbp]
\begin{center}
\includegraphics[width = 0.5\textwidth]{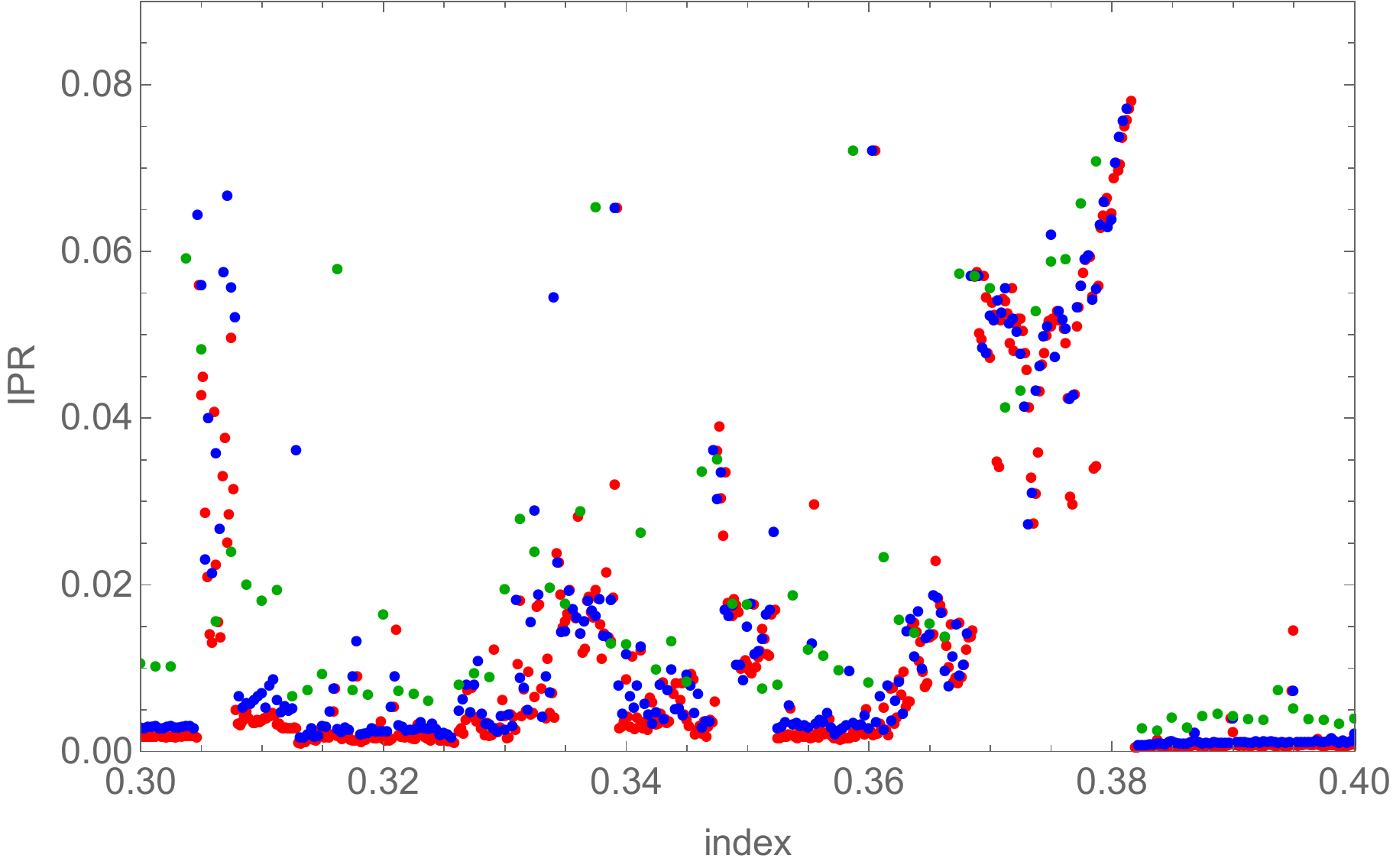}
\caption{IPR vs. eigenstate index for the unfrustrated model. $V = 1.33, p = 1.75$; $L = 800$ (green), $L = 3200$ (blue) and $L = 5600$ (red). The states between 0.36 and 0.38 appear to be largely localized.}
\label{detail1}
\end{center}
\end{figure}

\begin{figure}[htbp]
\begin{center}
\includegraphics[width = 0.5\textwidth]{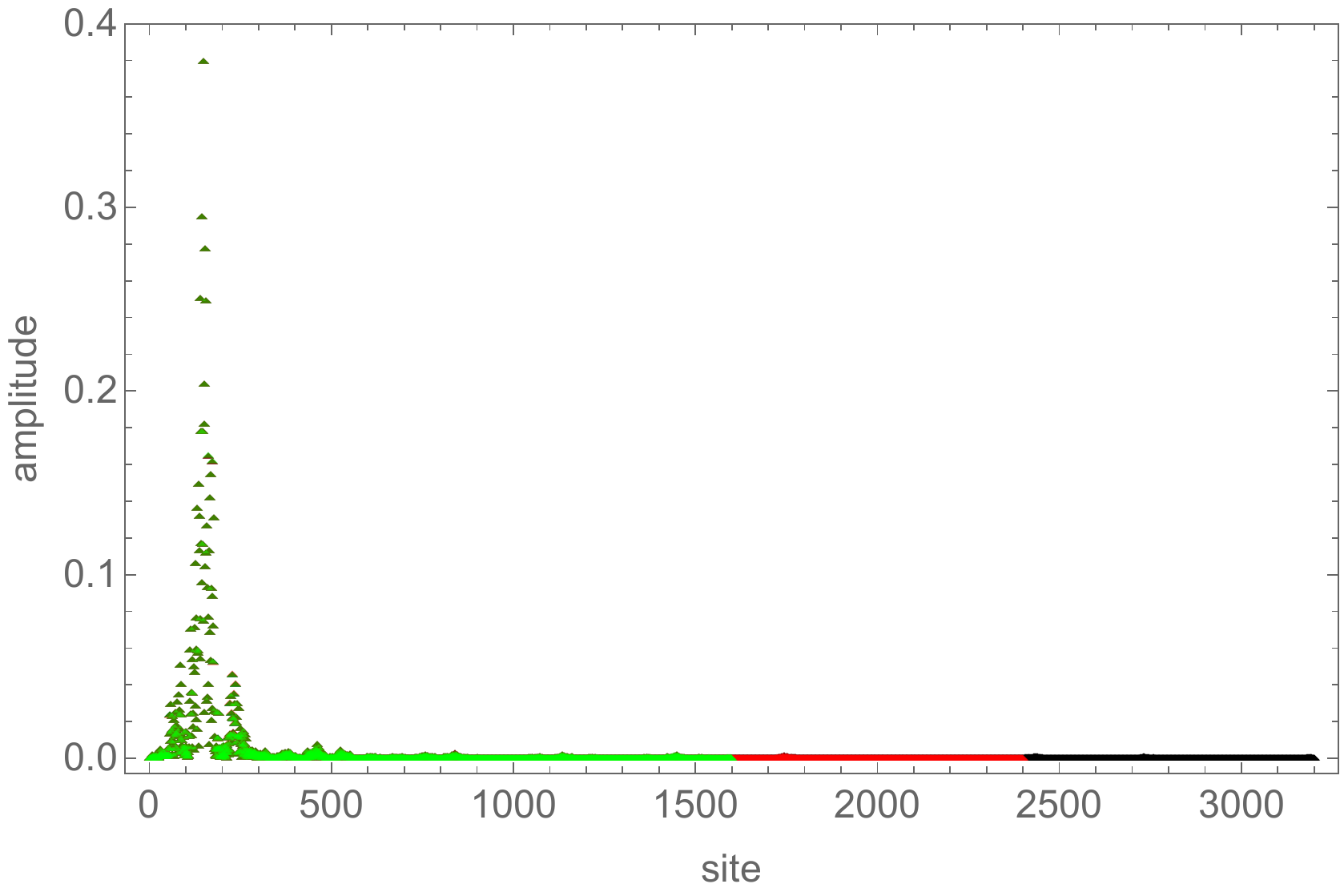}
\caption{Wavefunction of a localized eigenstate, for the same parameters as above, at various system sizes. The index of this state, counting from the top of the spectrum, is $999 (L = 1600), 1500 (L = 2400), 1998 (L = 3200)$. The overall phase $\phi = 0$ in Hamiltonian~\eqref{PLAAmodel}.}
\label{detail2}
\end{center}
\end{figure}

\begin{figure}[tbp]
\begin{center}
\includegraphics[width = 0.5\textwidth]{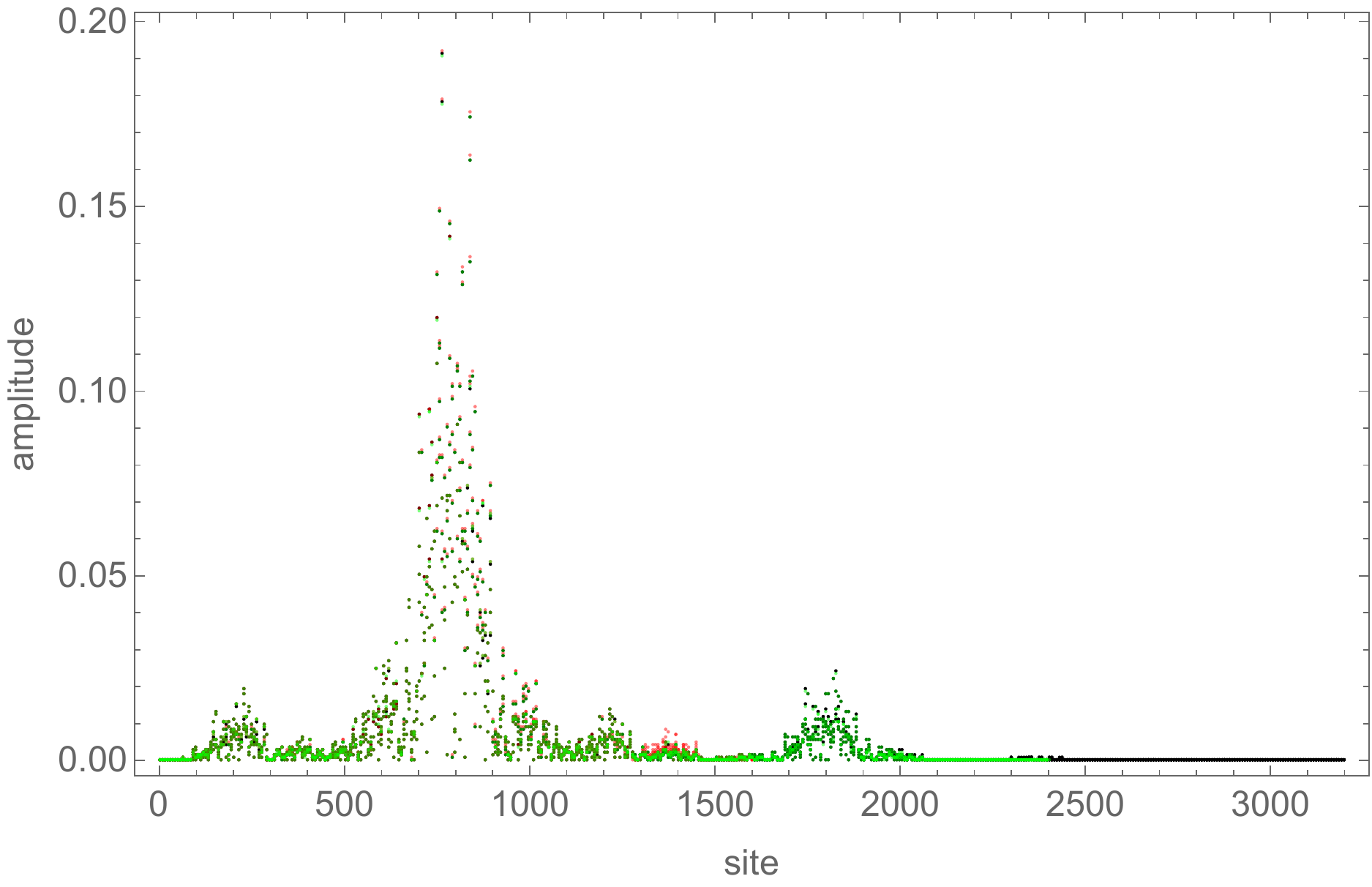}
\caption{Wavefunction of a localized eigenstate, for the same parameters as above, at various system sizes. The index of this state, counting from the top of the spectrum, is $674 (L = 1600), 1012 (L = 2400), 1348 (L = 3200)$. The overall phase $\phi = 0$ in Hamiltonian~\eqref{PLAAmodel}.}
\label{detail3}
\end{center}
\end{figure}

In this Appendix we provide a specific example to back up our claim that localized states emerge for $V < 2$ in the mixed phase. Specifically, Fig.~\ref{detail1} presents data on the scaling of IPR with system size for $V = 1.33, p = 1.75$. Evidently, there is a region of the spectrum for which the IPR does not appear to decrease with system size, indicating that localized states are present. Fig.~\ref{detail2} shows a particular localized wavefunction from this regime: the properties of this wavefunction are insensitive to system size. We estimate that for these parameters, 3-5$\%$ of the spectrum is localized, but determining the fraction accurately would require studying much larger system sizes. Fig.~\ref{detail3} shows an eigenstate that is neither clearly localized nor delocalized: it is spread out over $\sim 100$ sites, and its width is even larger; nevertheless, it is confined to a relatively small fraction of the system. 

\section{Size-dependence of heterogeneity}\label{appB}

In this section we discuss the finite-size scaling of the critical heterogeneity. This discussion is somewhat orthogonal to the main text: there, we identified the change in critical behavior by looking at the $p$-dependence of the critical heterogeneity at a fixed size. The heterogeneity evidently grows faster with system size in the mixed phase than at the critical point, and this difference in behavior allows us to identify the critical point. 

\begin{figure}[tbp]
\begin{center}
\includegraphics[width = 0.45\textwidth]{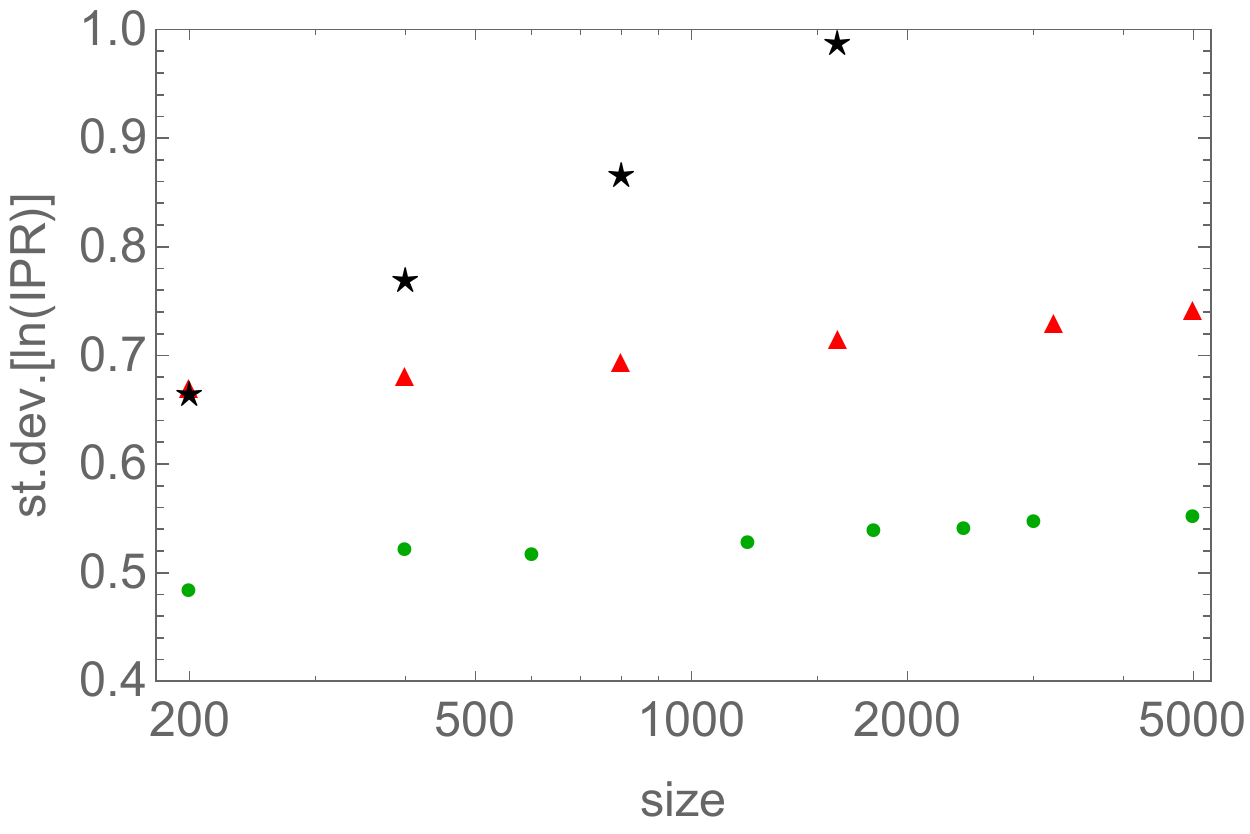}
\caption{Heterogeneity vs. system size for $p = 10$ at criticality (green circles), $p = 2$ at criticality (red triangles) and $p = 1.8, V = 3$ in the mixed phase (black stars). All results are averaged over fifty values of the phase $\phi$.}
\label{hetscaling}
\end{center}
\end{figure}

We first discuss a few limiting behaviors. For a conventional diffusive metal, the spatial profile of eigenfunctions is essentially random. A typical wavefunction occupies $o(N)$ sites, with standard deviation $o(\sqrt{N})$. This implies that the heterogeneity should decrease as $1 / \sqrt{N}$. Deep in the ballistic phase, one naively expects every site to occupy a finite fraction of the sites, although the precise fraction might vary from state to state. This would imply a saturating heterogeneity at large $N$. In a phase with both localized and delocalized wavefunctions, the distribution of $\ln(\ipr)$ should be bimodal, with a peak at a finite number and another at $\ln N$, giving a heterogeneity that scales as $\ln N$, which is indeed what we see (Fig.~\ref{hetscaling}). 

At the critical point, there are no localized states, yet the heterogeneity does seem to grow logarithmically with system size (albeit with a much weaker slope than in the mixed phase [Fig.~\ref{hetscaling}]). This behavior is possible, e.g., if the critical eigenstates are fractal, but different states have different fractal dimensions. The growth of the heterogeneity for $p = 2$ seems at least as fast as that for large $p$, indicating that the jump in the heterogeneity discussed in the main text survives in the thermodynamic limit.

\section{Structure of the low-frequency conductivity}\label{appC}

\begin{figure}[tbhp]
\begin{center}
\includegraphics[width=0.45\textwidth]{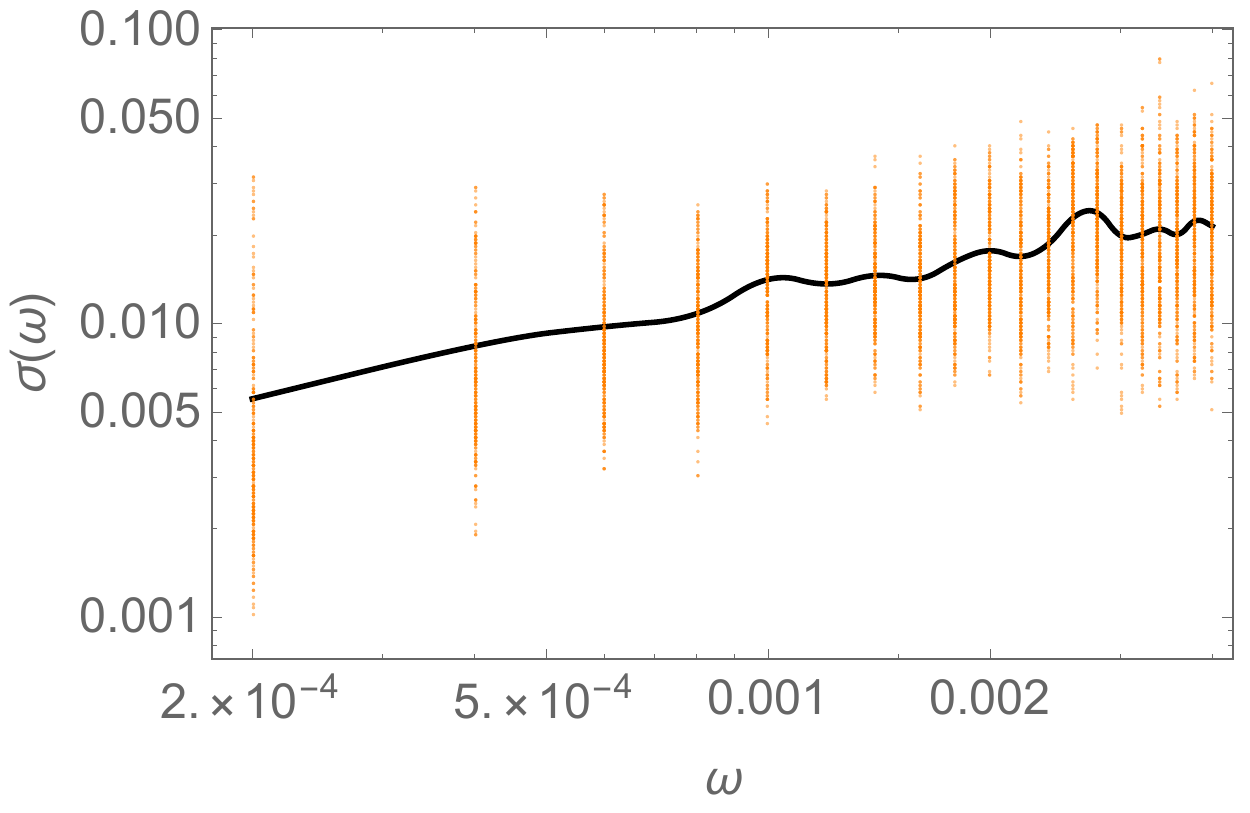}
\includegraphics[width=0.45\textwidth]{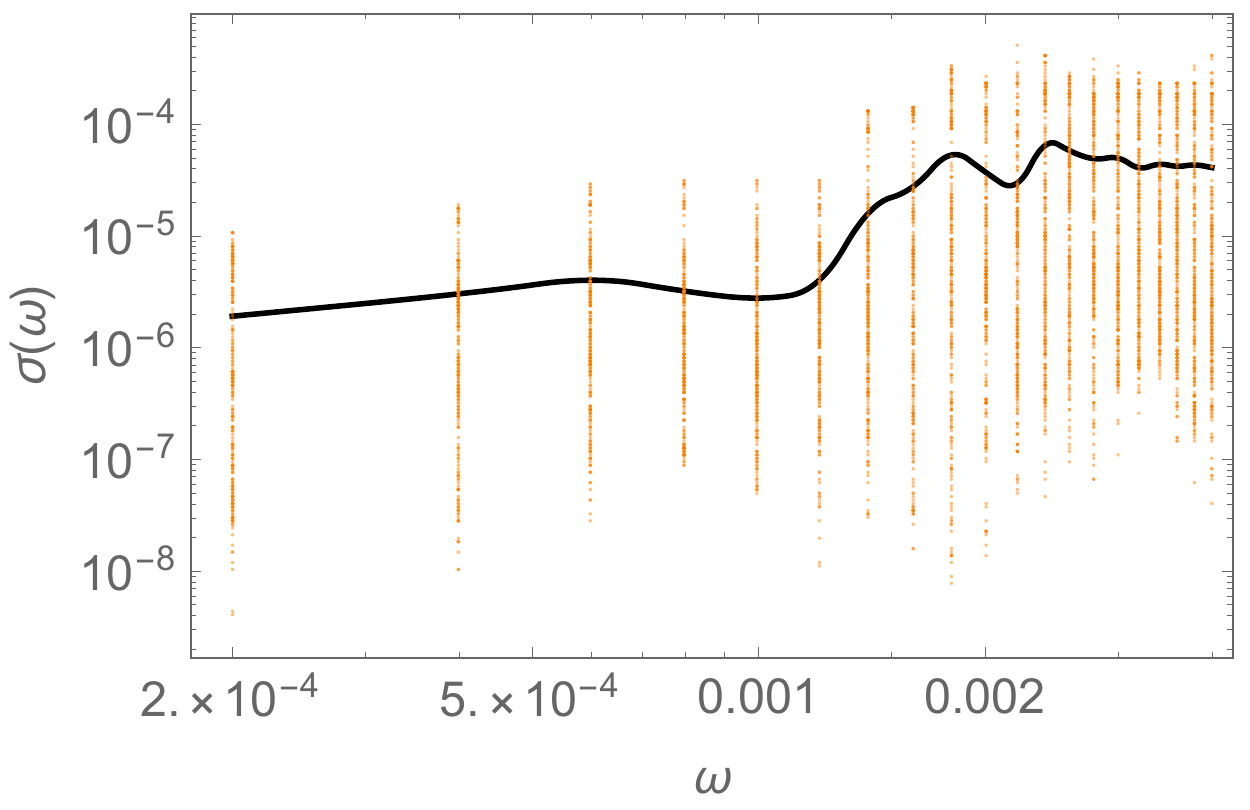}
\caption{Sample-to-sample fluctuations in the low-frequency conductivity, for $p = 2$ (upper panel) and $p = 4$ (lower panel). Black lines denote the sample-averaged conductivity; orange points correspond to single samples.}
\label{condflucts}
\end{center}
\end{figure}

In this Appendix we discuss the fluctuations of the low-frequency conductivity. We first summarize our approach for obtaining the numerical results in the main text: we numerically compute the quantity $\omega^2 x_{ij}^2 L_\eta(E_i - E_j - \omega)$, where $L_\eta(x) \equiv (1/\eta) 1/( + (x/\eta)^4)$ falls off slightly faster than a Lorentzian. (The rationale is to decrease the extent to which the tails of Lorentzians contaminate low-frequency data.) Summing over all pairs of states and normalizing this quantity by the Hilbert space size gives the optical conductivity. For our plots we chose $\eta = 10^{-4}$. 

Fig.~\ref{condflucts} shows the sample-to-sample fluctuations of the conductivity, which are manifestly large. These fluctuations are smaller for more weakly decaying power laws, as the system is more self-averaging in this case for reasons discussed in Sec.~\ref{conductivity}. For $p = 4$, sample-to-sample fluctuations range over three orders of magnitude; for $p = 2$, they are substantially weaker, at least on a logarithmic scale. The absolute extent of the fluctuations is notable: the system size used for these calculations is $L = 2000$, which is much larger than any intrinsic scale, and it is somewhat unexpected that samples should be so far from self-averaging at these sizes.

\section{Models with broken self-duality} \label{appD}

\begin{figure}[htb]
\begin{center}
\includegraphics[width=0.45\textwidth]{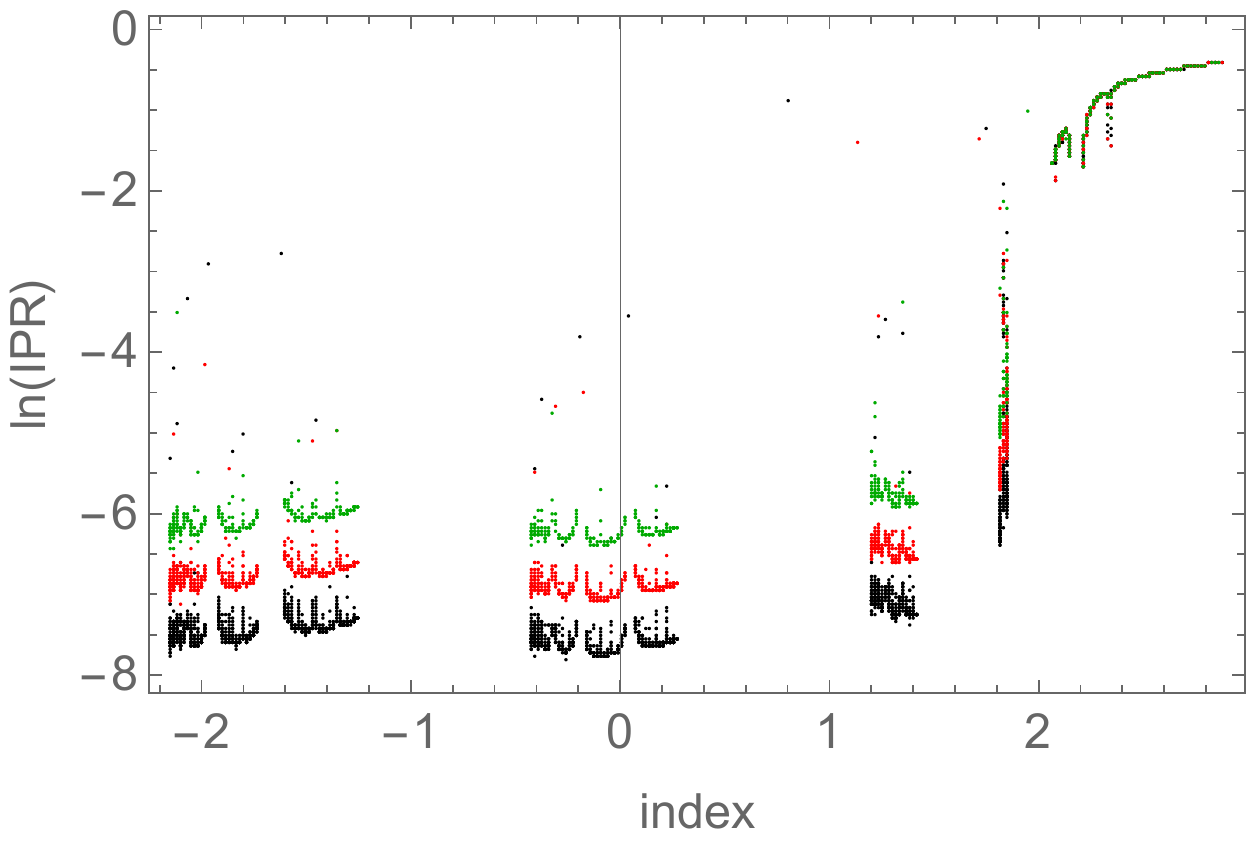}
\includegraphics[width=0.45\textwidth]{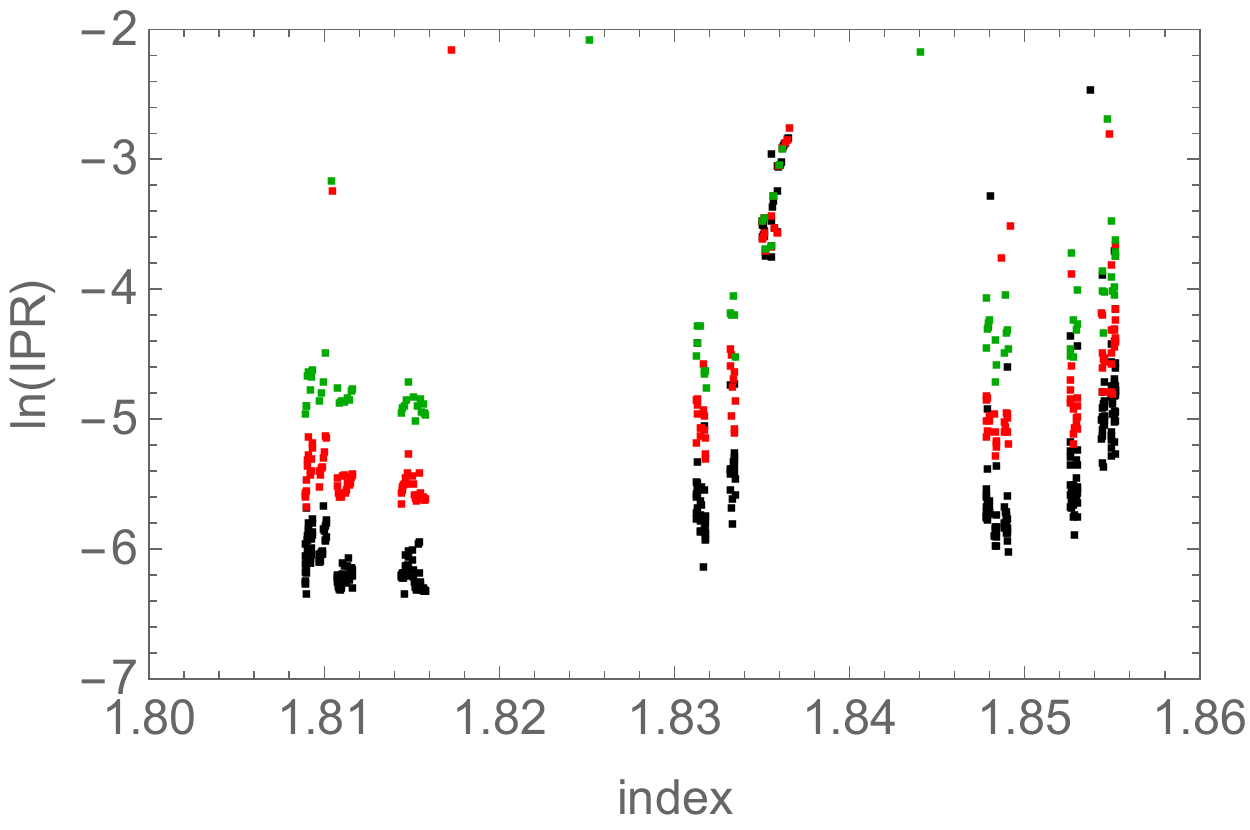}
\caption{IPR vs. eigenvalue for a model with power-law hopping ($p = 1.6$) but a cosine potential. Here, $V = 1$. The spectrum consists of clearly delocalized and clearly localized subbands, as well as one critical subband. The lower panel zooms in on the critical subband, and shows similar fine structure to that in Fig.~\ref{pd}.}
\label{psd}
\end{center}
\end{figure}

In practice, models with broken self-duality, featuring either power-law hopping or shaped potentials but not both, are simpler to realize experimentally. It is natural to ask which of the features discussed here survive in that limit. One of the implications of this work is that localized states \emph{should} be present in this model even for interactions $V < 2$ (because power-law hopping flattens parts of the dispersion, favoring localization). A second implication is that---because the critical states are \emph{spectrally} distinct from either localized or delocalized states---there should not be a simple mobility edge in such models as there is in conventional Anderson insulators. Rather, each subband should either be entirely localized or delocalized, or be itself critical (and thus have additional fine structure). 

These implications are borne out numerically (Fig.~\ref{psd}). There are generically both delocalized and localized subbands. Generally each subband is fully localized or delocalized. However, in narrow ranges of the coupling, a subband can also be critical (Fig.~\ref{psd}), and these critical subbands have fine structure that is very similar to what we see in the mixed phase of the self-dual model. 

%


\end{document}